\documentclass[preprint2,times,tighten]{aastex6}
\pdfoutput=1 
\usepackage[figure,figure*]{hypcap}

\usepackage{graphicx}
\graphicspath{{fig/}}

\def\uo{\textrm{u\kern -1.1ex\raisebox{0.6ex}{$^\circ$}}} 

\usepackage[encapsulated]{CJK}
\usepackage{ucs}
\usepackage[utf8x]{inputenc}
\newcommand{\cntext}[1]{\begin{CJK}{UTF8}{gbsn}#1\end{CJK}}

\newcommand{\skipthis}[1]{}

\newcommand{\lsim}{${\raisebox{-.9ex}{$\stackrel{\textstyle<}{\sim}$}}$ }

\def\nh2d{$\rm{NH_2D}$}

\def\nh3{$\rm{NH_3}$}
\def\NH3{$\rm{NH_3}$}

\def\n2hp{$\rm{N_2H^+}$}
\def\NTH{$\rm{N_2H^+}$}

\def\h2o{$\rm{H_2O}$}
\def\h2{$\rm{H_2}$}

\def\hcop{$\rm{HCO^+}$}

\def\msun{\,$M_\odot$}

\def\um{\,$\mu\mathrm{m}$}

\def\kms{\,km~s$^{-1}$}
\def\cm2{\,$\rm{cm^{-2}}$}
\def\cm3{\,$\rm{cm^{-3}}$}
\def\cms{\,$\rm{cm^{-2}}$}
\def\cmc{\,$\rm{cm^{-3}}$}

\def\vlsr{$V\rm{_{LSR}}$}

\def\11{(1,1)}
\def\22{(2,2)}
\def\33{(3,3)}
\def\44{(4,4)}
\def\55{(5,5)}
\def\t21{$T_{21}$}
\def\r31{$R_{31}$}

\def\ga{G28.34+0.06}

\def\spt{\emph{Spitzer}}
\def\her{\emph{Herschel}}

\let \amp = \&

\shorttitle{Automated filaments identification using minimum spanning tree}
\shortauthors{Wang et al.}
\submitted{\textcolor{blue}{ApJS in press. Submitted 2016 May 8, accepted 2016 July 8.}}

\begin{document}

\title{A census of large-scale ($\ge$10 pc), velocity-coherent, dense filaments in the northern Galactic plane: automated identification using minimum spanning tree}
\author{Ke Wang \cntext{(王科)}\altaffilmark{1}}
\author{Leonardo Testi \altaffilmark{1,2,3}}
\author{Andreas Burkert \altaffilmark{4,5}}
\author{C. Malcolm Walmsley \altaffilmark{6,7}}
\author{Henrik Beuther \altaffilmark{8}}
\author{Thomas Henning \altaffilmark{8}}
\email{kwang@eso.org}
\altaffiltext{1}{European Southern Observatory (ESO) Headquarters,
Karl-Schwarzschild-Str. 2,
85748 Garching bei M\"{u}nchen,
Germany}
\altaffiltext{2}{Excellence Cluster Universe, Boltzmannstr. 2, 85748 Garching bei M\"{u}nchen, Germany}
\altaffiltext{3}{INAF -- Osservatorio astrofisico di Arcetri, Largo E. Fermi 5, 50125 Firenze, Italy}
\altaffiltext{4}{University Observatory Munich, Scheinerstrasse 1, D-81679 Munich, Germany}
\altaffiltext{5}{Max-Planck-Institute for Extraterrestrial Physics, Giessenbachstrasse 1, D-85758 Garching, Germany}
\altaffiltext{6}{INAF -- Osservatorio astrofisico di Arcetri, Largo E. Fermi 5, 50125 Firenze, Italy}
\altaffiltext{7}{Dublin Institute of Advanced Studies, Fitzwilliam Place 31, Dublin 2, Ireland}
\altaffiltext{8}{Max-Planck Institute f\"{u}r Astronomie, K\"{o}nigstuhl 17, D-69117 Heidelberg, Germany}


\begin{abstract}

Large-scale gaseous filaments with length up to the order of 100 pc are on the upper end of the filamentary hierarchy of the Galactic interstellar medium.
Their association with respect to the Galactic structure and their role in Galactic star formation are of great interest from both observational and theoretical point of view.
Previous ``by-eye'' searches, combined together, have started to uncover the Galactic distribution of large filaments, yet inherent bias and small sample size limit conclusive statistical results to be drawn.
Here, we present (1) a new, automated method to identify large-scale velocity-coherent dense filaments, and (2) the first statistics and the Galactic distribution of these filaments.
We use a customized minimum spanning tree algorithm to identify filaments
by connecting voxels in the position-position-velocity space, using the Bolocam Galactic Plane Survey spectroscopic catalog.
In the range of $7.^{\circ}5 \le l \le 194^{\circ}$,
we have identified 54 large-scale filaments and derived 
mass ($\sim 10^3 - 10^5$\msun), length (10--276 pc), linear mass density (54--8625 \msun\,pc$^{-1}$), aspect ratio, linearity, velocity gradient, temperature, fragmentation, Galactic location and orientation angle.
The filaments concentrate along major spiral arms.
They are widely distributed across the Galactic disk, with 50\% located within $\pm$20 pc from the Galactic mid-plane and 27\% run in the center of spiral arms.
An order of 1\% of the molecular ISM is confined in large filaments.
Massive star formation is more favorable in large filaments compared to elsewhere.
This is the first comprehensive catalog of large filaments useful for a quantitative comparison with spiral structures and numerical simulations.

\end{abstract}

\keywords{catalogues, stars: formation, ISM: clouds, ISM: structure, Galaxy: structure}
\maketitle

\section{Introduction} \label{sec:intro}
The interstellar medium (ISM) has a highly filamentary and hierarchical structure. On the upper end of this filamentary hierarchy are large-scale gaseous filaments with length up to the order of 100 pc. What is their distribution in our Galaxy and what role do they play in the context of Galactic star formation? Answers to these questions are important for a critical comparison with theoretical studies and numerical simulations of galaxy formation and filamentary cloud formation. The observational key to answer these questions is a homogeneous sample of large filaments across the Galaxy identified in a uniform way. 

Studies in the past years have revealed a number of large filaments with a wide range of aspect ratios and morphologies, from linear filaments to a collection of cloud complexes.
\cite{Goodman2014} find that the 80 pc long infrared dark cloud (IRDC) ``\object{Nessie}'' \citep{Jackson2010} in the southern sky can be traced up to 430 pc in the position-position-velocity (PPV) space in $^{12}$CO (1--0), guided by connecting the IR-dark patches presumably caused by high column density regions extincting the otherwise smooth IR background emission from the Galactic plane. They argue that the Nessie runs in the center of the Scutum-Centaurus spiral arm in the PPV space, so termed as a ``bone'' of the Milky Way.
In a follow up study, \cite{Zucker2015} searched the region covered by the
MIPSGAL \citep[\spt/MIPS Galactic Plane Survey, $|l|<62^\circ, \, |b|<1^\circ$;][]{MIPSGAL} focusing on the PPV loci of arms expected by various spiral arm models, finding 10 bone candidates with length 13--52 pc and aspect ratio 25--150.
\cite{Ragan2014-GFL} and \cite{Abreu2016} extend this ``mid-IR extinction'' method to a blind search, i.e., not restricting to arm loci but the full extend of the observed PPV space. They find 7 and 9 filaments with length 38--234 pc in part of the first 
and fourth Galactic quadrants covered by the GRS (Galactic Ring Survey; \citealt{Jackson2006-GRS}), and the ThrUMMS (Three-mm Ultimate Mopra Milky Way Survey; \citealt{Barnes2015-ThrUMMS}), respectively. The aspect ratios of those filaments are not well defined due to the complex morphology, but inferring from the figures in the papers, the typical aspect ratio is much less than 10.

In contrast to the indirect\footnote{Indirect because ``IR-dark'' does not necessarily correspond to a dense cloud; it can be caused by a real ``hole in the sky'' \citep{Jackson2008,Wilcock2012}.}
``mid-IR extinction'' method, \cite{me15} identify large filaments directly based on emission at far-IR wavelengths near the spectral energy distribution (SED) peak of cold filaments. They develop a Fourier Transform filter to separate high-contrast filaments from the low-contrast background/foreground emission. Fitting the SED built up from the multi-wavelength \her\ data from the Hi-GAL survey \citep{Hi-GAL}, they derive temperature and column density maps, and have used those maps to
select the ``largest, coldest and densest'' filaments. They present 9 filaments with length 37--99 pc and aspect ratio 19--80, identified primarily from the GRS field.

These systematic searches have started to uncover the spatial distribution of large filaments in our Galaxy, revealing filaments within and outside major spiral arms. However, with different searching methods and selection criteria, in addition to inherent bias from manual inspection, it is difficult to cross compare the results from these studies. The small sample size also limits the robustness of statistical attempts \citep[e.g., see discussion in][]{me15}.
All the above mentioned searches start from a ``by-eye'' inspection of dust features (either mid-IR extinction or far-IR emission), identify candidate filaments, and then verify the coherence in radial velocity using gas tracers --- spectral line data.

We automate the identification process by applying a customized minimum spanning tree algorithm to the 
PPV space. We present the first homogeneous sample of 54 large-scale velocity-coherent filaments in the range of $7.^{\circ}5 \le l \le 194^{\circ}$
(see exact coverage in \autoref{sec:data}).
We derive mass, length, linearity, aspect ratio, velocity gradient and dispersion, temperature, column/volume density, fragmentation, Galactic location and orientation angle. For the first time, we are able to investigate the Galactic distribution of their physical properties, and to estimate the fraction of the ISM confined in large filaments and star formation therein.

We describe the data set in \autoref{sec:data} and present our identification method in \autoref{sec:method}.
The identified sample of filaments and their physical properties and statistics are presented in \autoref{sec:para}, followed by a discussion of the nature and implication of the filaments in \autoref{sec:discuss}. Main conclusions are summarized in \autoref{sec:sum}.
Following the spirit of \cite{me15}, we focus on the densest filaments traced by millimeter dust continuum emission, and not the more diffused CO filaments.

\section{Data: a complete spectroscopic catalog} \label{sec:data}

The Bolocam Galactic Plane Survey (BGPS) is a blind mapping of the northern Galactic plane at 1.1 mm using the Caltech Submillimeter Observatory 10m telescope with an effective resolution of $33''$, revealing over 8400 continuum sources \citep{Aguirre2011_BGPS1,Rosolowsky2010_BGPS2}. Spectroscopic follow-ups carried out by \cite{Schlingman2011} and \cite{Shirley2013} have observed all the 6194 BGPS sources in the longitude range of $7.^{\circ}5 \le l \le 194^{\circ}$ in dense gas tracers $\rm{HCO^+}$ (3--2) and $\rm{N_2H^+}$ (3--2), using the 10m Heinrich Hertz Submillimeter telescope, with a FWHM beamwidth of $15''$. The detection rate is about 50\%, and about 99\% of the detections show a unique velocity component.

From these observations \cite{Shirley2013} compiled a complete spectroscopic catalog of 3126 sources with a single velocity component resolved in $\rm{HCO^+}$ (3--2) and/or $\rm{N_2H^+}$ (3--2). In a typical temperature range of 10--20 K of the BGPS sources \citep{Dunham2011-BGPS-NH3}, these two lines have a critical density of $\sim 10^6$ \cmc\ and
an effective excitation density of the order $10^4 - 10^5$ \cmc\ \citep{Shirley2015}, thus they trace very dense gas.
In the following, we refer to these sources as ``dense BGPS sources''.
At a typical distance of few kpc, a detection of the lines towards a BGPS 1.1 mm continuum peak marks the presence of pc scale dense gas, and a chain of such clumps connected in PPV means a rather prominent structure.
Therefore, this catalog is an excellent data set for searching for velocity-coherent filaments. 

Note that the coverage of the BGPS spectroscopic catalog is contiguous in the range of $7^{\circ}.5 \le l \le 90^{\circ}.5$, $|b| \le 0^{\circ}.5$, with latitude coverage flaring up to $|b| \le 1^{\circ}.5$ in several longitude cuts. In the outer Galaxy, four selected regions were observed
($l$ range in [98.85, 100], [110, 112], [132.5, 138.5], and [187.5, 193.5] degrees).
The BGPS spectroscopic survey \citep{Shirley2013} used the version 1 BGPS continuum source catalog \citep{Aguirre2011_BGPS1}. In the release of the version 2 catalog, \cite{Ginsburg2013_BGPSv2} resolved an offset in flux scale:
$S_{\rm v2} = 1.5 S_{\rm v1}$. In this study, we use the flux from v2 (\autoref{sec:para}).

\begin{figure}
\includegraphics[width=.5\textwidth,angle=0]{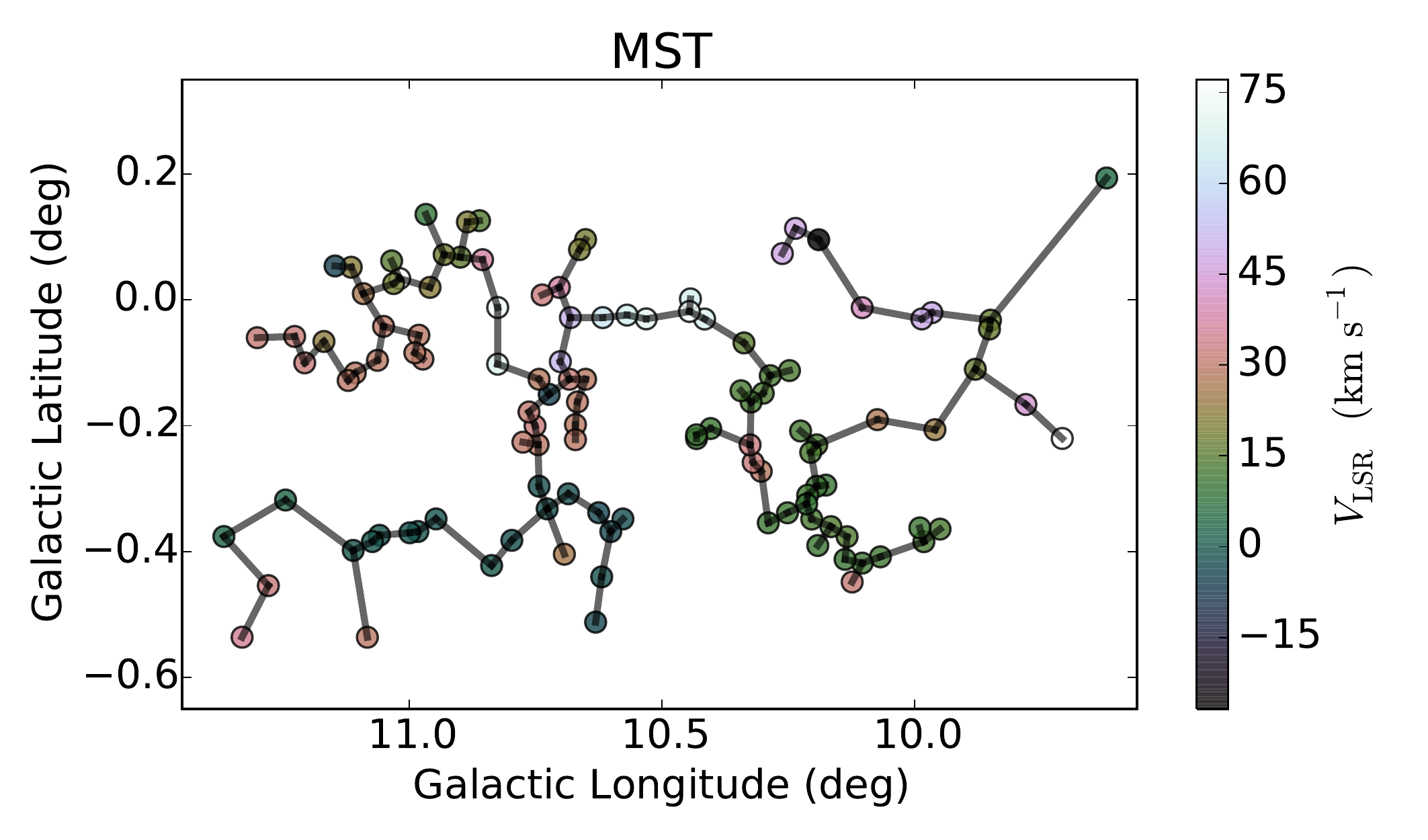}
\includegraphics[width=.5\textwidth,angle=0]{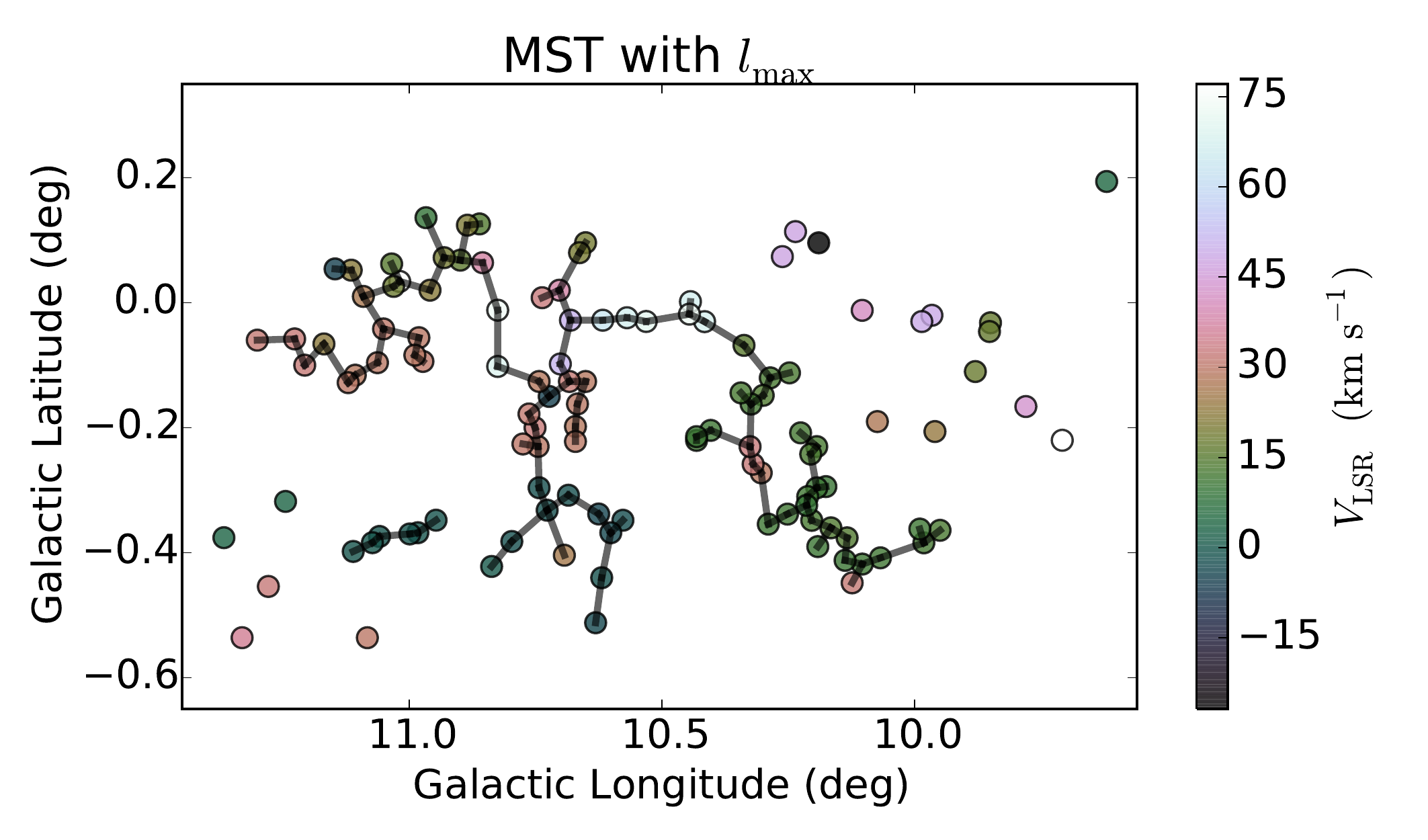}
\includegraphics[width=.5\textwidth,angle=0]{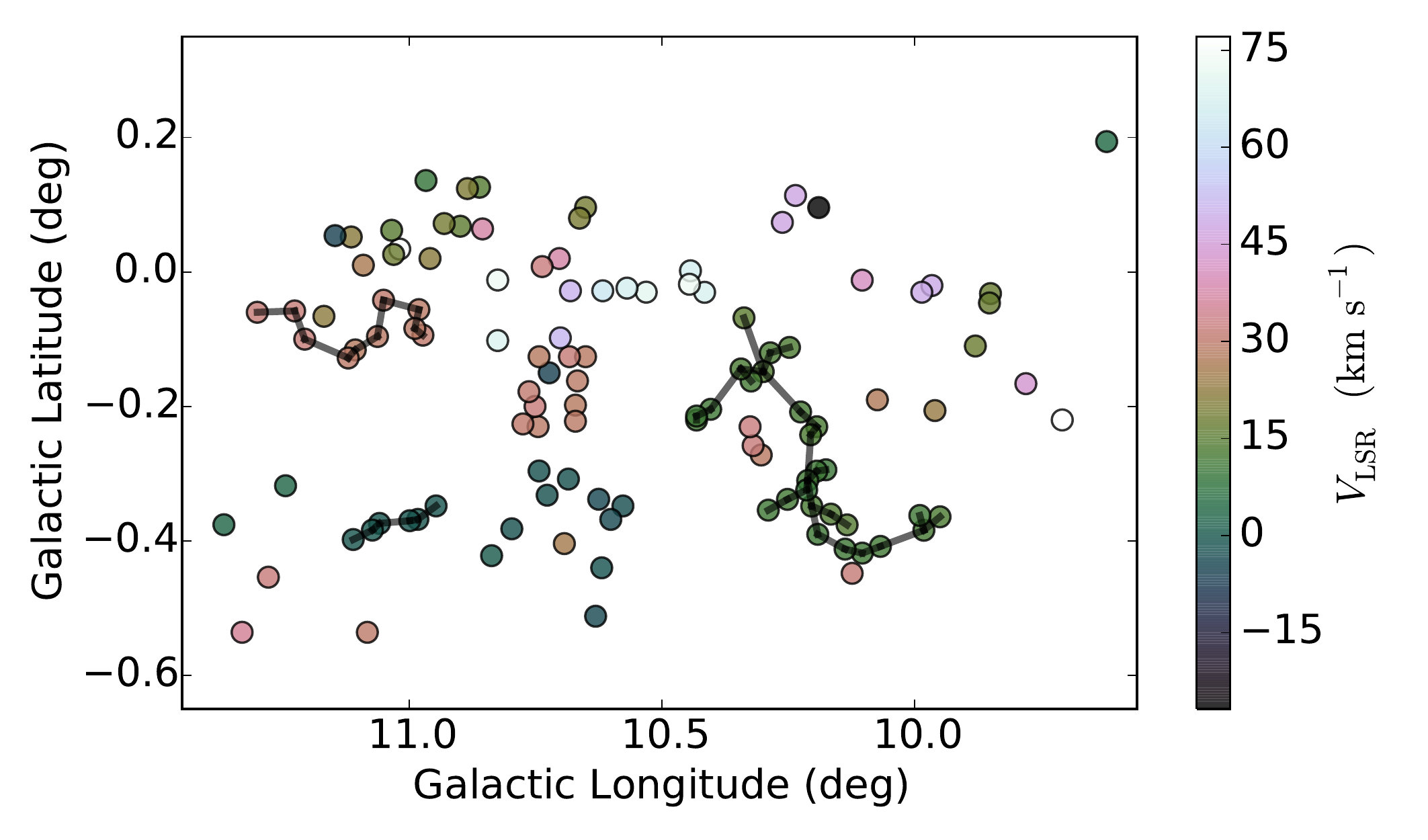}
\caption{
Demonstration of the customized MST applied to a subset of the Galactic longitude.
\textit{Upper}: a MST.
\textit{Middle}: MSTs with a maximum edge length and minimum number of clumps (criteria 1--2 in \autoref{sec:method}).
\textit{Lower}: MSTs with edge limited to $\Delta v <2$\kms\ (criteria 1--3).
Circles represent BGPS sources color coded by radial velocity as shown in the color bar.
}
\label{fig:demo}
\end{figure}

\section{Method: connecting dots using customized MST} \label{sec:method}

The minimum spanning tree (MST) was first introduced by 
Otakar Bor{\uo}vka
to optimize the cost of electrical grids by minimizing its total length. It is now widely used in general optimization of a variety of networks (see review by \citealt{MST2001}).
We adopt the MST algorithm to isolate coherent filaments out of a PPV catalog bearing in mind that ``coherence'' means ``close proximity'' in position \textit{and} velocity.
Our method is demonstrated in \autoref{fig:demo}.
A MST connects all the nodes (BGPS clumps) in a graph with the cost of minimum sum of edge lengths, where ``edge'' refers to the separation between two nodes (\autoref{fig:demo}a). 
We customize the MST such that a graph is connected not in one MST, but in a collection of MSTs with the following criteria:
\begin{enumerate}
\item[(1)]
The accepted MST must contain at least 5 BGPS clumps: $N_{\rm cl} \ge 5$.
\item[(2)]
Only edges shorter than a maximum length can be connected ($\Delta L <0.1^{\circ}$, \autoref{fig:demo}b).
\item[(3)]
For any two clumps to be connected, the difference in line-of-sight velocity ($\Delta v$) must be less than 2 \kms\, (\autoref{fig:demo}c).
\end{enumerate}
The maximum edge length and velocity difference (criteria 2--3) are chosen based on characteristics of previously known filaments \citep{me15}.
As in \cite{me15}, we have used the ``\object{Snake}'' nebula, one of the first identified IRDCs \citep{carey98}, as a primary guide to test criteria (2--3). In addition to its prominent IR extinction feature, the Snake is known to have velocity-coherent structures at multiple spatial scales \citep{carey2000,me14,my-Springer-sum}. Together, these properties make the Snake a good test case.
During the tests, we relax criteria (2--3) by increasing $\Delta L$ and $\Delta v$ until the MST starts to (incorrectly) connect unrelated BGPS sources into the Snake. We also require the algorithm to be able to identify as many as possible previously known filaments from other studies (see \autoref{sec:knownFL}), but not to connect unrelated sources.
Criterion (1) is arbitrary, but we note that there is no difference if we set the minimum number of clumps $N_{\rm cl}$ as 5 or 6. It is clear in \autoref{tab:fl} which filaments would be picked up should we increase the required $N_{\rm cl}$.

So far, these MSTs are \textit{coherent} structures in the $(l,b,v)$ space, but not necessarily \textit{filamentary} structures. To further select filaments we introduce a linearity check.
For a given filament, we fit a straight line to the $(l,b)$ data points.
The fitted line represents the filament's major axis.
Linearity is defined as the ratio between 
the spread (standard deviation) of data points measured along the filament's major axis to the spread along minor axis: $f_L = {\sigma_{\rm major}}/{\sigma_{\rm minor}}$.
After visual inspection we accept structures with 
\begin{enumerate}
\item[(4)] Linearity $f_L >1.5$
\end{enumerate}
as filaments.
Note that the linearity is defined to quantify
how far the structure is away from a linear structure. For a straight line, $f_L \rightarrow \infty$. 
For bent or wiggling filaments, which is often the case, the linearity is much smaller than the aspect ratio. For instance, the famous IRDC ``Snake'' (F7 in \autoref{tab:fl}) has an aspect ratio of 43, while its linearity is only 4 because of its sine wiggling in the $(l,b)$ space \citep{me14,me15,my-Springer-g11}.
Finally, after determining the distance (\autoref{sec:dist}) we accept long filaments with
\begin{enumerate}
\item[(5)] Projected length $\ge$10 pc
\end{enumerate}
as large-scale filaments for the interest of this study.

Applying the methodology to the BGPS spectroscopic catalog, we have identified 91 velocity-coherent structures (satisfying criteria 1-3), 70 of which are linear filaments (satisfying criteria 1-4). Among these, 54 are large-scale filaments (satisfying criteria 1-5), including 48 in the first Galactic quadrant where the BGPS coverage is contiguous, and 5/1 in the second/third quadrants respectively, where the BGPS coverage is targeted to known star formation regions.
Of the 54 filaments, only 6 are previously known
(F7, F13, F25, F33, F36, and F41; \autoref{sec:knownFL}).
Clearly, our filaments identification method depends on free parameters (like many other methods) \footnote{Other filament-finding methods include: getfilaments \citep{soft:getfilaments-Menshchikov2013}, DisPerSE \citep{soft:DisPerSE-Sousbie2011,soft-FilTER-Panopoulou2014,LiGX2016-FL}, FilFinder \citep{soft:FilFinder-Koch2015}, Hessian matrix \citep{Schisano2014,Salji2015-FL-Orion}, and Bisous model \citep{soft-Tempel2016-Bisous}. 
Our customized MST and the Bisous model work on discrete points, and others work on continuous images.}, hence the identified filaments and their properties depend on the chosen parameters of criteria (1--5). Here we have chosen reasonable criteria to select the \textit{representative} large-scale velocity-coherent filaments based on previously known filaments.

Each of the 54 filaments are plotted in \autoref{fig:rgb} in a two-color view, where the mid-IR 24 or 22\um\ emission \citep{MIPSGAL,WISE} is shown in cyan and the (sub)millimeter 0.87 or 1.1 mm emission \citep{Csengeri2016-ATLASGAL,Ginsburg2013_BGPSv2} is shown in red. The MST edges are drawn to outline the filaments.
Most (40 out of the 54) filaments are IRDCs, while 14 are IR-bright filaments.
The filaments show a wide range of filamentary morphologies (see next section).

\section{Physical parameters} \label{sec:para}

\autoref{tab:fl} lists physical parameters of the 54 identified filaments. 
Column (1) assigns identification numbers running from F1 to F54.
Col. (2--4) flux weighted longitude, latitude (in degree), and local standard of rest (LSR) velocity (\kms). For instance, for a filament containing $n$ clumps, its flux weighted longitude is
$l_{\rm wt} = \frac{\sum_{i=1}^{n} F_i\times l_i}{\sum_{i=1}^{n} F_i}$, where $F_i$ and $l_i$ are the BGPS flux and the longitude of the $i$th clump, respectively.
Col. (5--6) distance (kpc) and its type (see \autoref{sec:dist}).
Col. (7) number of clumps in the filament.
Col. (8--9) length of the filament in degree and pc by summing all the edges in the filament.
Col. (10) velocity gradient (\kms\,pc$^{-1}$): the mean of all the edges.
We caution that both velocity and length are subject to projection effect.
Col. (11) dispersion of the central velocity of all the clumps in the filament (\kms).
Col. (12--13) minimum and maximum temperature of the clumps (see \autoref{sec:T}).
Col. (14-15) filament mass (in unit $10^3$\msun) and linear mass density (\msun/pc).
Mass is computed from the integrated BGPS v2 1.1\,mm dust emission flux measured in a polygon encompassing the filament guided by the MST, adopting the \cite{Ossenkopf1994} dust opacity law and $\beta = 1.5$, and accounting for the different temperatures in the clumps (\autoref{sec:T}).
Col. (16--17) molecular hydrogen column density ($10^{22}$ \cms) and volume density ($10^3$ \cmc) of the filament. These are estimated by simplifying the filament as a cylinder with a length of the filament length and a diameter of the mean major axes of the clumps.
Col. (18) aspect ratio $f_A$ estimated by dividing filament length with the averaged major axes of the clumps.
Col. (19) linearity $f_L$ (see definition in \autoref{sec:method}).
Col. (20) $R_{\rm gc}$, Galactocentric radius (kpc).
Col. (21) $z$, vertical distance (pc) to the physical Galactic mid-plane
after correction for the Sun's displacement of 25 pc above the mid-plane and the true position of the Galactic Center \citep{Goodman2014,me15}.
Col. (22) $\theta$, orientation angle (degree) between the filament's long axis and the physical Galactic mid-plane. Positive/negative angle means Galactic latitude increases/decreases with increasing longitude (\autoref{fig:demo}c).
Col. (23) Morphology class as defined in \cite{me15}:
L: linear straight or L-shape;
C: bent C-shape;
S: quasi-sinusoidal shape;
X: crossing of multiple filaments;
H: head-tail or hub-filament system.
Some filaments are characterized by more than one class.
Different morphologies may have resulted from different filament formation processes. For example, expansion of bubbles can produce C shaped filaments, collision of bubbles can make S shaped filaments, gravitational contraction of a clump embedded in a sheet can make H type filaments, while turbulence and converging flows can make filaments of any shape.
Col. (24--27) Galactic coordinate boundary of the filament (degree).

The last rows of \autoref{tab:fl} list statistics of these physical parameters.
These include minimum, maximum, median, mean, standard deviation, skewness ($K$), and kurtosis ($K$).
Skewness is a measure of symmetry. A symmetric distribution has $S = 0$, while negative/positive value of $S$ means asymmetric tails with lower/higher values around the mean, respectively.
Kurtosis is a measure of how the distribution is compared to a normal distribution (e.g., Gaussian): a Gaussian distribution has $K=0$, while a negative or positive value $K$ means the distribution is a flatter or more centrally peaked distribution than Gaussian, respectively.
These statistics provide a description of each parameter, which is discussed in \autoref{sec:stat}.

The major sources of uncertainty of the parameters originate from the uncertainties in distance, dust opacity, and projection. Because projection is unknown, we do not correct for it in \autoref{tab:fl}, but we discuss its effect on different parameters here.
The dust opacity tabulated in \cite{Ossenkopf1994} is subject to a factor of 2 uncertainty.
The typical distance uncertainty is 10-30\%, depending on different distance types (see \autoref{sec:dist}).
Distance uncertainty propagates to other parameters. In the following estimation we use a conservative  30\% distance uncertainty.
\textit{Length}:
includes distance uncertainty of 30\% and projection. Due to projection, the length is a lower limit. For a random inclination angle ($\phi$, defined as the angle between line-of-sight and filament's long axis) of 1 radian, $L_{\rm obs} = L \times {\rm sin} (\phi) = 0.84 L$.
\textit{Velocity gradient}:
includes distance uncertainty of 30\% and projection.
Projection affects both velocity and length, in a form of tan($\phi$), thus projection can lead to an over- or underestimation of the velocity gradient.
For $\phi = 30^{\circ}, 57.3^{\circ}, 75^{\circ}$,
the observed velocity gradient is the true value times cos($\phi$)/sin($\phi$) = 1.73, 0.64, 0.27, respectively.
\textit{Mass}:
includes uncertainties in dust opacity, distance (scaling as $d^2$), and flux (10\%, \citealt{Ginsburg2013_BGPSv2}).
Together, these uncertainties propagate to $<$50\% uncertainty in mass.
Note we have accounted the temperature variance across the filament \citep[similar to][]{me15}, an improvement over other studies in which a uniform temperature is \textit{assumed} for all filaments \citep[e.g.][]{Ragan2014-GFL,Abreu2016}.
\textit{Mass per unit length}:
Compared to mass, $M/L$ has a smaller dependence on distance uncertainty (scaling as $d^1$), thus the error is less than that of mass.
But $M/L$ is affected by projection in the form of $1/L$.
\textit{Column density}:
defined as mass/(length$\times$width), the distance error is eliminated, thus the uncertainty in column density is less than that of mass.
Projection affects length but not width, so projection is in the form of $1/L$.
Additionally, simplifying filaments into cylinders can cause uncertainties for some filaments which deviate from a linear structure (with a relatively small linearity $f_L$).
\textit{Volume density}:
defined as mass/(length$\times$width$\times$depth), where depth is assumed to be the same as width.
Compared to mass, volume density has a smaller dependence on distance uncertainty (scaling as $1/d$), thus the uncertainty is less than that of mass.
But volume density is affected by projection in the form of $1/L$.
Similar to column density, simplifying filaments into cylinders can cause additional uncertainties for some filaments which deviate from a linear structure.

\subsection{Distance estimation} \label{sec:dist}
Distance is determined by three methods, listed in decreasing order of robustness:
(1) type P: trigonometric parallax measurements of associated masers from the BeSSeL project \citep[][Sanna, private communication]{BeSSeL,Brunthaler2009,Moscadelli2009,XuY2011,Immer2013,Wu2014-BeSSeL}.
(2) type ML: maximum likelihood distance from Bayesian evaluation of kinematic distance using external data to place priors \citep{Ellsworth2015,Ellsworth2013-dist}.
(3) type KN, KF:
near or far kinematic distance computed using the procedure of \cite{Reid2009} with updated Galactic parameters from \cite{Reid2014}.

Wherever available, we use parallax distance, then Bayesian distance, then kinematic distance.
Of the 54 filaments, 8 are assigned for type P, 39 for ML, 4 for KN, and 3 for KF distances.
When more than one clump in a given filament has ML distance, the median of the clump distances are used, computed after excluding extreme values. This occurs, for example, when a filament is generally IR-dark (therefore more likely to be located at near distance), but a minority of its clumps are IR-bright and so the far distance is assigned by the ML evaluation.
Kinematic distance is used in 8 cases. The distance ambiguity is resolved with the IR emission/extinction (see \autoref{fig:rgb}) \textit{and} the fact that all clumps in a given filament should have a consistent distance computed by the \cite{Reid2009} code.

\subsection{Temperature estimation} \label{sec:T}
We evaluate the temperature of every clump from three resources, listed in order of decreasing priority to search for a match:
(1) gas kinetic temperature determined from \nh3 for a subset of the BGPS sample \citep{Dunham2011-BGPS-NH3}, 
(2) same but for a subset of the ATLASGAL (Atacama Pathfinder EXperiment Telescope Large Area Survey of the Galaxy) sample \citep{Wienen2012-NH3}, and
(3) dust color temperature determined by comparing 350\um\ and 1.1\,mm fluxes for a subset of BGPS sample \citep{Merello2015-BGPS-350um}. We exclude color temperatures with large uncertainties ($>$100\%).

When querying for \nh3 temperature the clump must match in position and velocity, and for color temperature only position is available for matching.
If a clump does not have a match, we assume a temperature of 15 K based on the average value of BGPS clumps \citep{Dunham2011-BGPS-NH3}.
Among the 54 filaments, 46 have at least one temperature match.
The minimum and maximum temperature values are listed in \autoref{tab:fl}.

\begin{figure*}
\centering
\includegraphics[width=.245\textwidth]{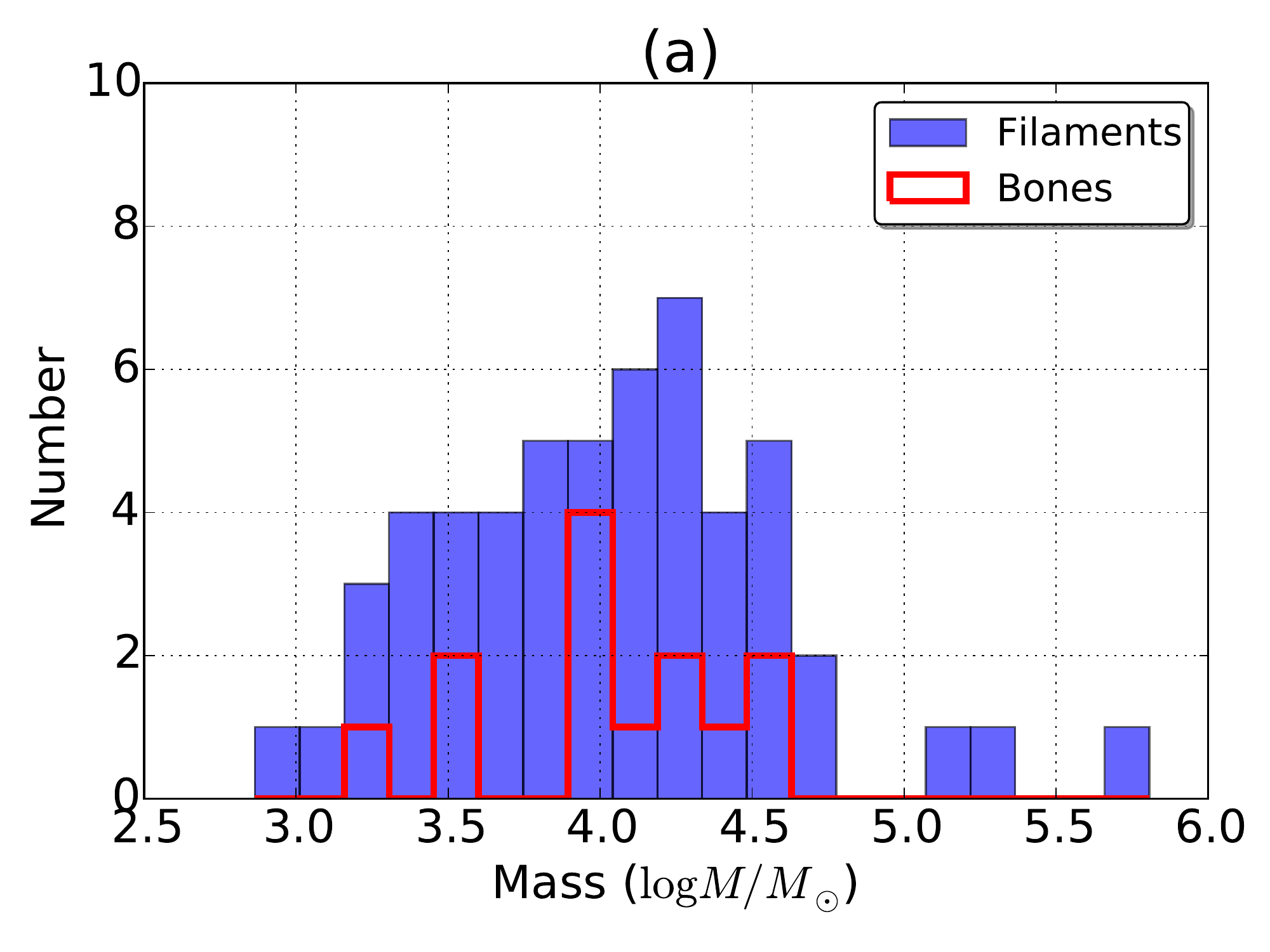}
\includegraphics[width=.245\textwidth]{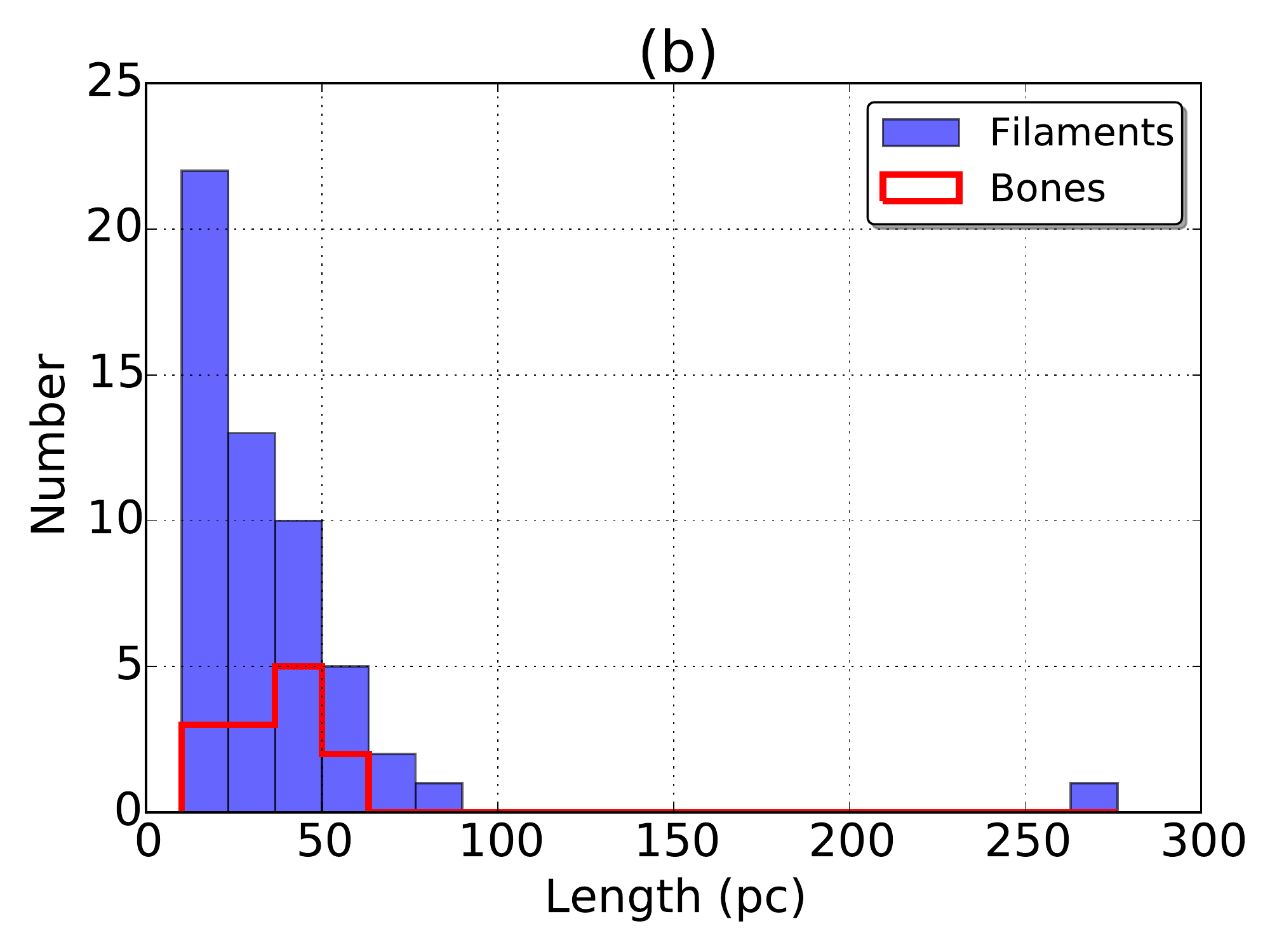}
\includegraphics[width=.245\textwidth]{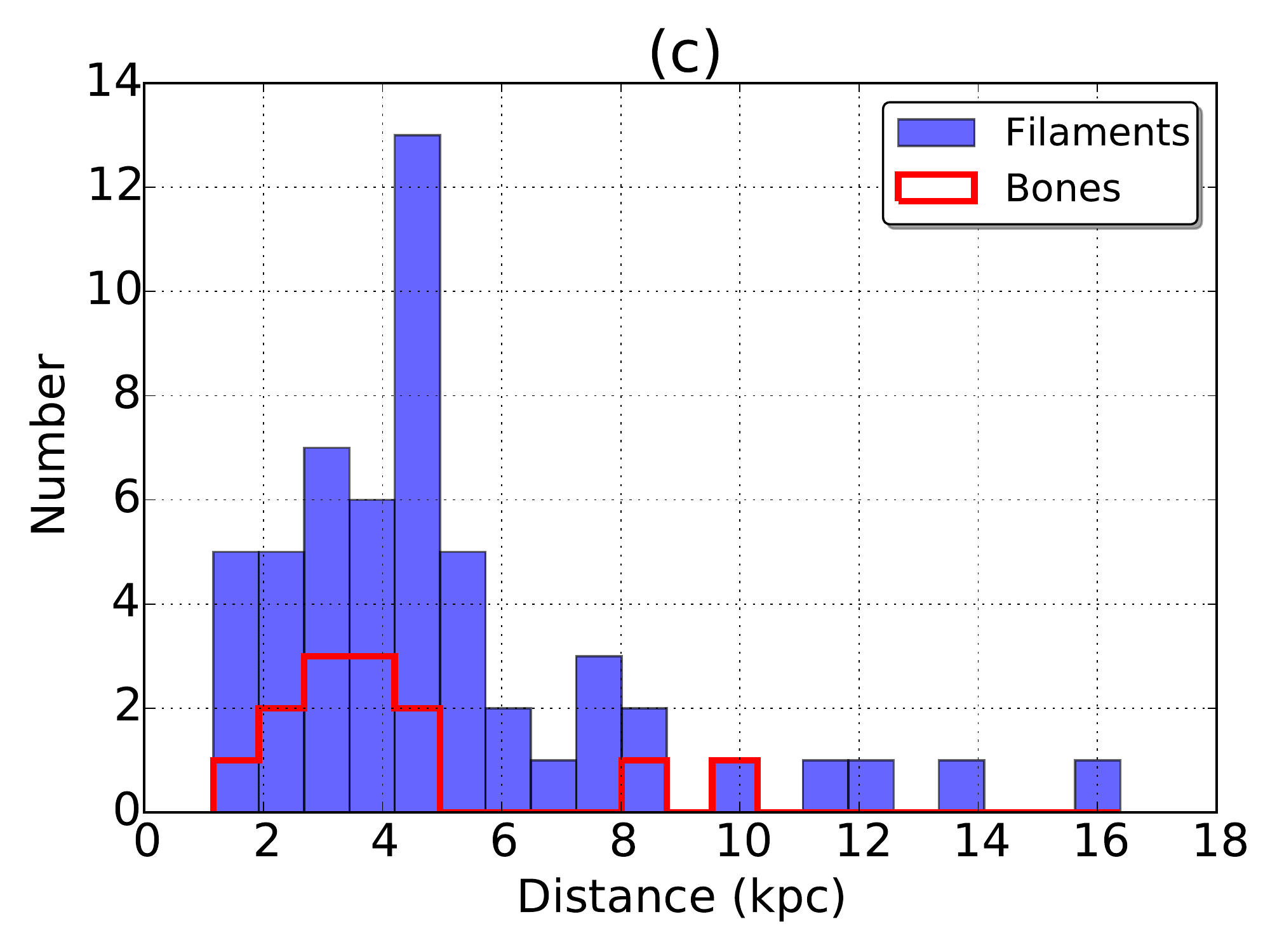}
\includegraphics[width=.245\textwidth]{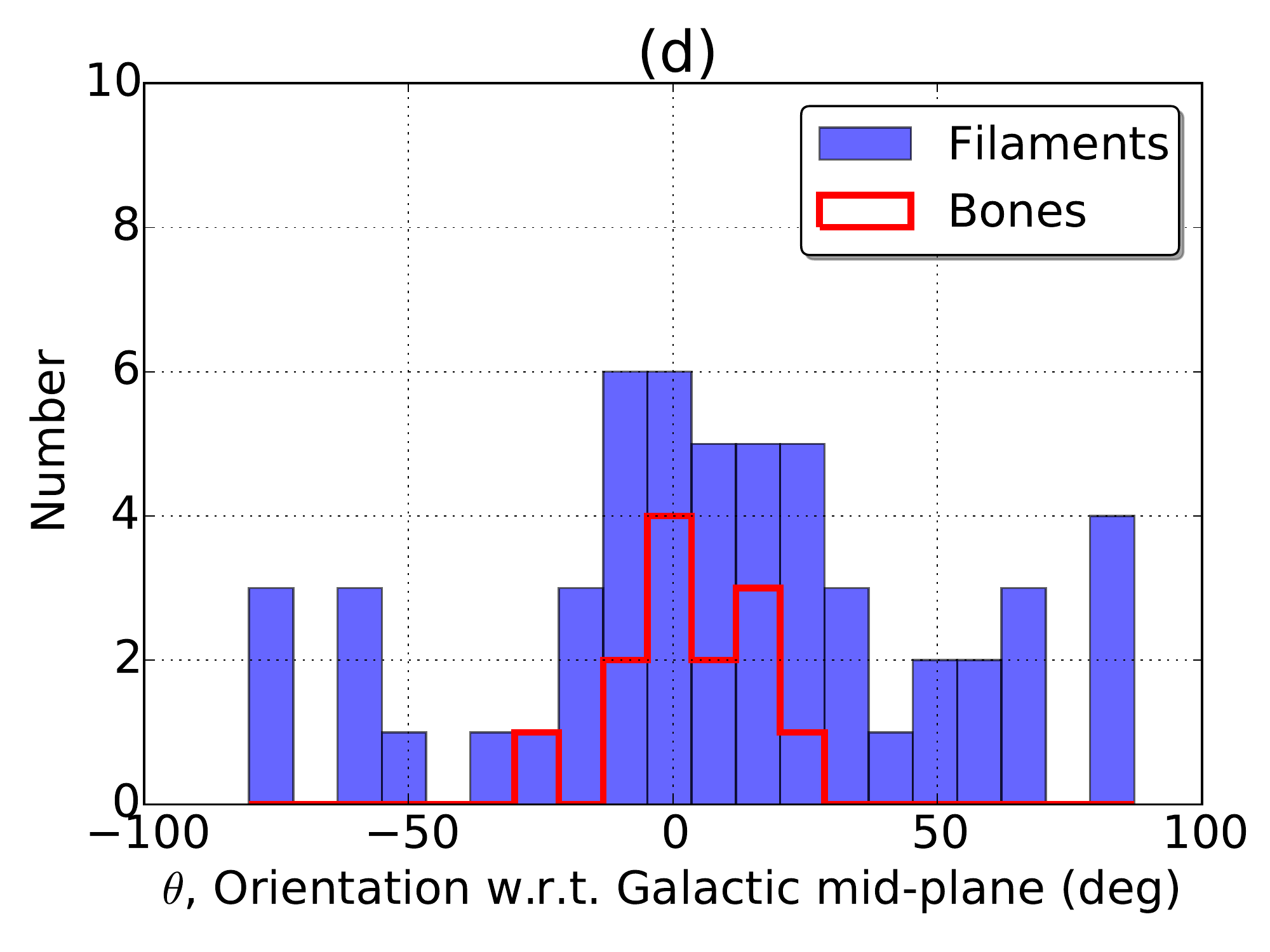}
\\
\includegraphics[width=.245\textwidth]{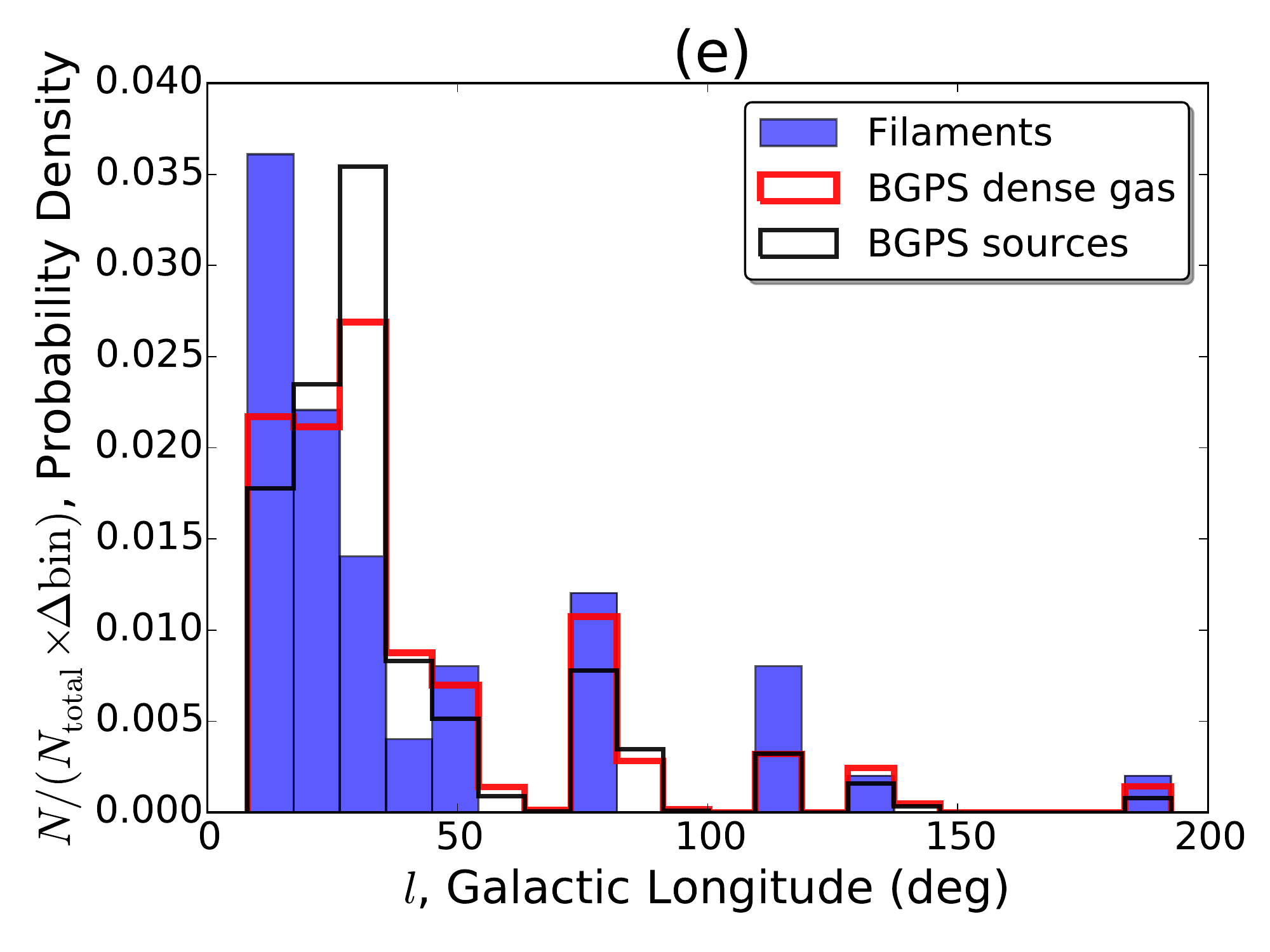}
\includegraphics[width=.245\textwidth]{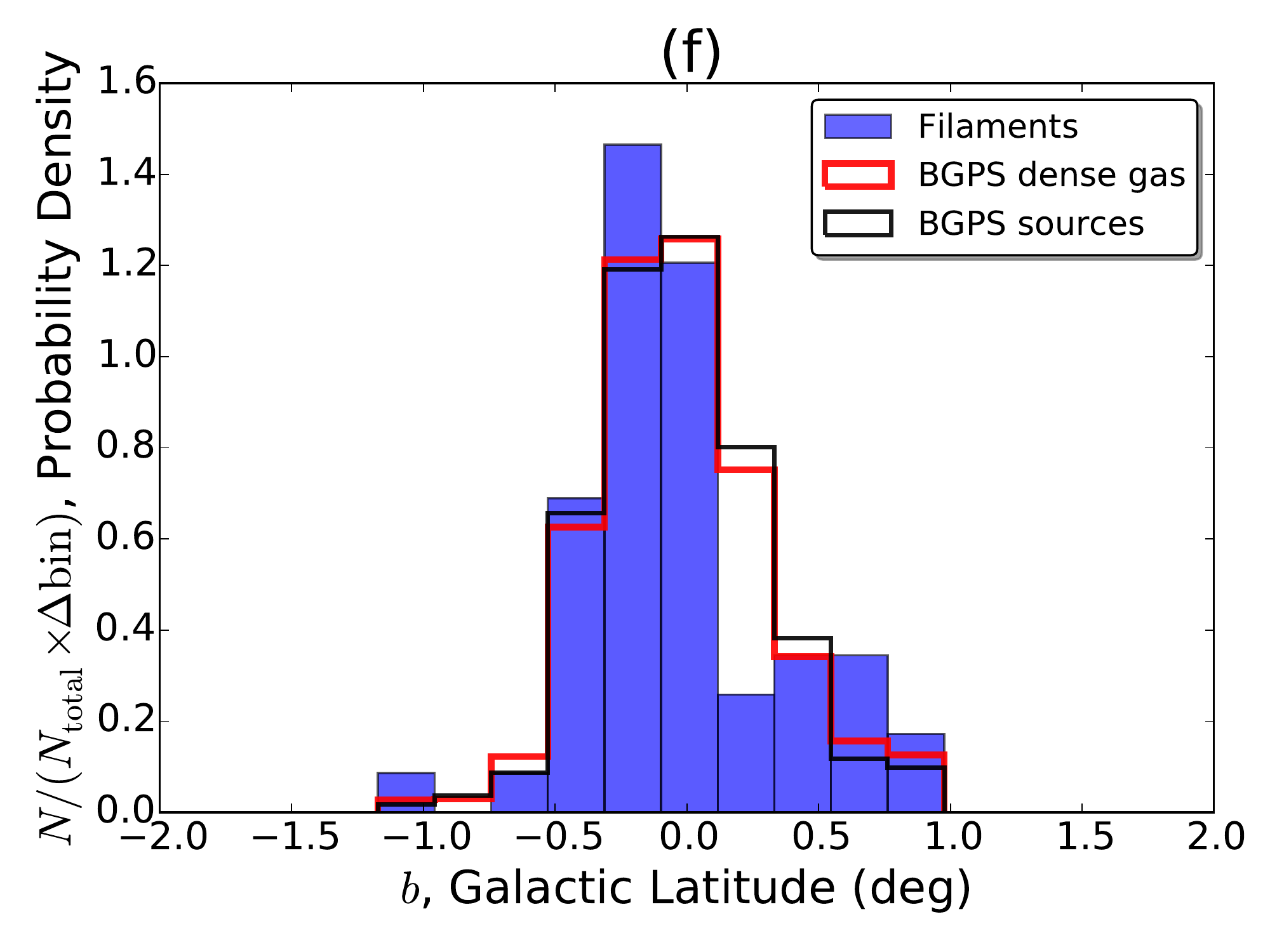}
\includegraphics[width=.245\textwidth]{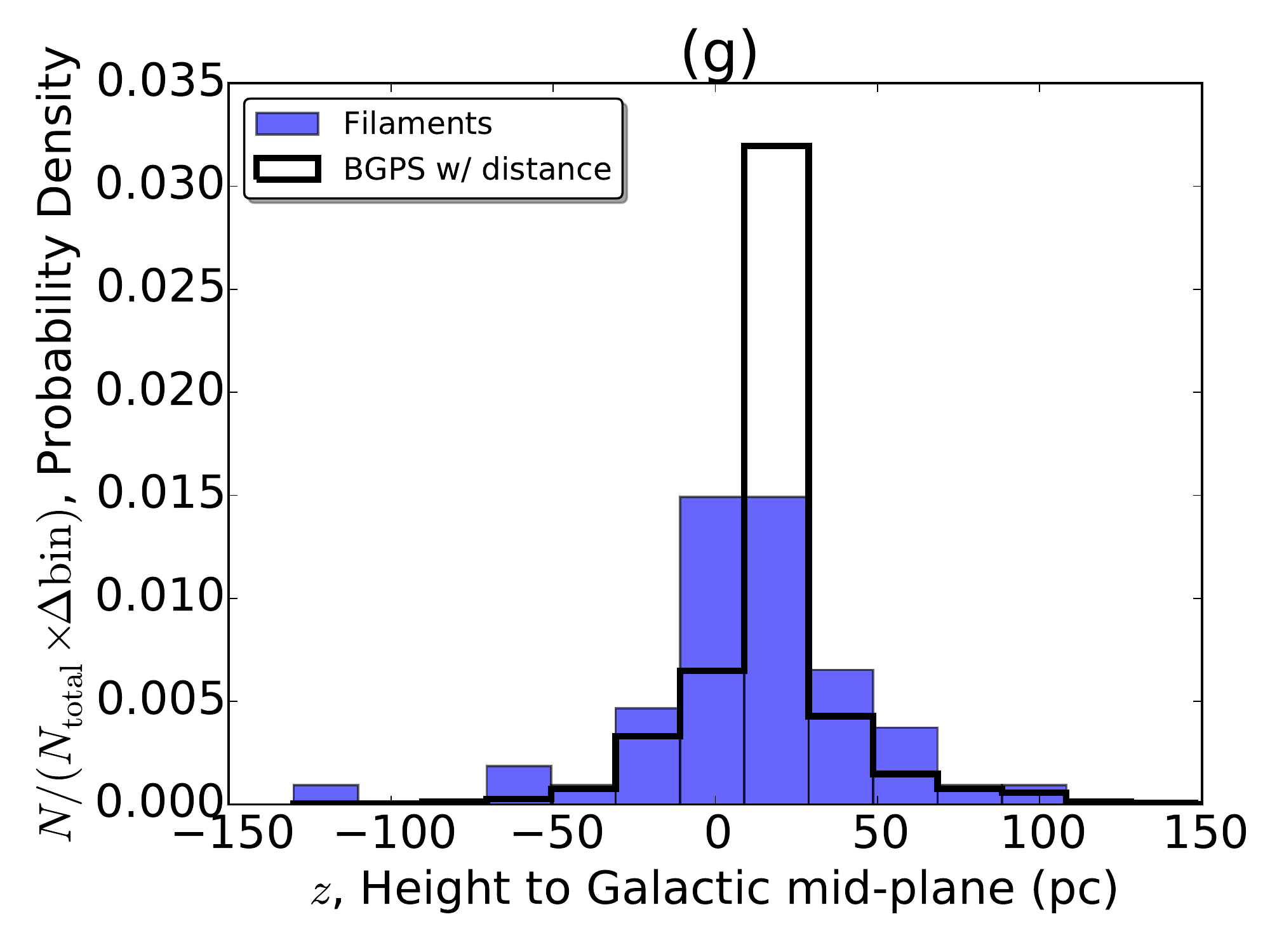}
\includegraphics[width=.245\textwidth]{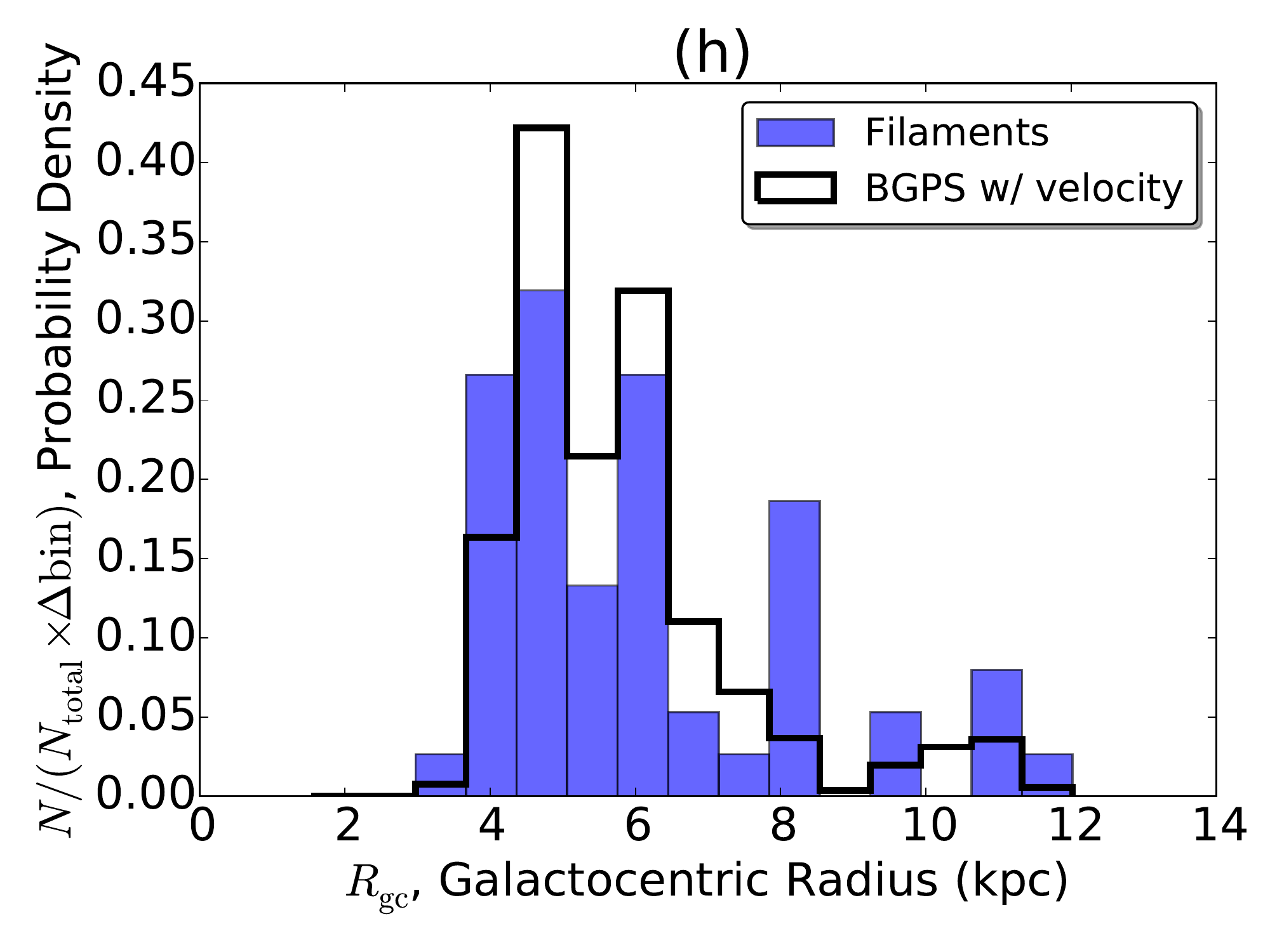}
\caption{
Panels (a--d) show histograms of filament mass, length, distance, and orientation angle. The 54 filaments and 13 bones are labeled.
Panels (e--h) plot normalized distribution of $l, b, z$ and $R_{\rm gc}$, comparing filaments with BGPS sources.
``BGPS sources'' are the BGPS clumps covered in the searching field.
``BGPS dense gas'' means the 3126 BGPS clumps with dense gas detection in \cite{Schlingman2011} and \cite{Shirley2013}.
The 1710 ``BGPS w/ distance'' source and 3508 ``BGPS w/ velocity'' sources are from \cite{Ellsworth2015}.
}
\label{fig:hist}
\end{figure*}

\subsection{Statistics of the parameters} \label{sec:stat}

\autoref{fig:hist} presents histograms of selected parameters. Panels (a-d) show mass, length, distance, and orientation angle for the filaments. Panels (e-h) present normalized histograms of $l, \, b, \, z, \, R_{\rm gc}$ where the distribution of filaments is compared to an appropriate BGPS sub-sample when data is available.

Among the 54 identified filaments, F5 stands out as an extreme: 276 pc and $6.4\times10^5$ \msun. Excluding this extreme, the other 53 filaments extend 10--93 pc over the projected sky, with a mass in the order of $10^3 - 10^5$ \msun. For all the 54 filaments, the angular length lies in the range of $0.17^{\circ}-1.50^{\circ}$ with a median of $0.35^{\circ}$. 
The column density $N_{\rm H_2}$ lies in the range of $(0.2-16.9) \times 10^{22}$ \cms\ with a median value of $0.9 \times 10^{22}$ \cms, and volume density $n_{\rm H_2}$ is in the range $(0.4-22) \times 10^3$ \cmc\ with a median value of $1.8 \times 10^3$ \cmc.
The filaments are 1--2 orders of magnitude denser than the filaments identified by the ``mid-IR extinction'' \citep{Ragan2014-GFL,Zucker2015,Abreu2016}, while as dense as the filaments identified from \her\ far-IR emission by \cite{me15}.

The mass-length relation is best fitted as ${\rm lg}L = 0.41\times {\rm lg}M - 0.19$. This suggests that the filaments are not $N$ small unrelated filaments that appear to be accidentally connected. Suppose the sub-units have a typical length $l$ and a typical mass $m$. The total length of the filament would then be $L=N\times l$ with mass $M=N\times m$, resulting $L \sim M$ and not $L \sim M^{0.41}$ as observed here.
The mass-length relation implies $M \sim L^{2.4}$, i.e., a fractal dimension of 2.4. This is comparable with the observed GRS molecular cloud catalog (although of whole clouds) in the Milky Way \citep{Roman2010-GRS} and with numerical simulations of supersonic gas turbulence \citep{Federrath2009}. The mass-length scaling of filaments, from large to small-scales, will be discussed in detail in a forthcoming paper (Burkert et al. in prep).

The orientation of the filaments is clearly not random. The distribution of the orientation angles $\theta$ is close to Gaussian ($K = -0.3$), showing a weak but significant concentration towards $0 ^{\circ}$.
The vertical distribution $z$ of the filaments is surprisingly not symmetric with respect to the Galactic plane ($S=-0.7$). It skews towards the negative values, with a mean at a positive value of $z=11.5$ pc. This is likely an observational bias (see \autoref{sec:bias}). Nevertheless, $z$ concentrates towards small values, with 50\% (27/54) filaments located within $|z| \le 20$ pc and 70\% (38/54) within  $|z| \le 30$ pc. Towards higher vertical positions the number of filaments decreases much faster than a Gaussian function ($K=2.8$).
Three parameters have a Gaussian-like distribution: the Galactocentric radius $R_{\rm gc}$ ($K=0.0$), the clump velocity dispersion $\sigma (v)$ ($K=-0.01$), and the flux weighted LSR velocity $v_{\rm wt}$ ($K=-0.3$). Notably, the mean velocity gradient along the filament is small ($0.43\pm0.31$ \kms\,pc$^{-1}$), but in broad agreement with simulations \citep{Duarte2016}.

Across the Galaxy, in general, the distribution of filaments follows the number density of BGPS sources, i.e., it is more likely to find a filament where the BGPS sources are crowed. Specifically, the probability density\footnote{The probability density is defined as
$P = N_{\rm bin}/(N_{\rm total} \times \Delta_{\rm bin})$, so that the integral of the histogram will sum to 1.}
of filaments ($P_{\rm Filament}$) largely agrees with the probability density of BGPS sources: $P_{\rm Filament} \sim P_{\rm BGPS}$, as binned in $l, \, b, \, z, \, R_{\rm gc}$ [\autoref{fig:hist}(a-d)]. 
This is expected because we used the BGPS spectroscopic catalog as input, which is a homogeneous subsample of the BGPS catalog. However, there are interesting exceptions.
First, a significantly lower filament probability ($P_{\rm Filament} \lsim 0.5\times P_{\rm BGPS} $)  is seen towards $l \sim 30^{\circ}$, the Scutum tangent. The same discrepancy is seen towards $b \sim 0.2^{\circ}$, $z \sim 20$ pc, and $R_{\rm gc} \sim 5$ kpc.
Second, on the other hand, a much higher $P_{\rm Filament}$ is found in the inner Galaxy of $b \sim 10^{\circ}$, towards Galactocentric radius of $R_{\rm gc} \sim 4$ and 8 kpc, and most significantly, towards zero Galactic vertical scale height, $z$. 
Third, the averaged $b$ and $z$ for filaments are closer to zero than that of BGPS sources. In another word, the filaments show a more symmetric distribution (than BGPS sources) with respect to the physical Galactic mid-plane.

\begin{figure}
\centering
\includegraphics[width=.49\textwidth,angle=0]{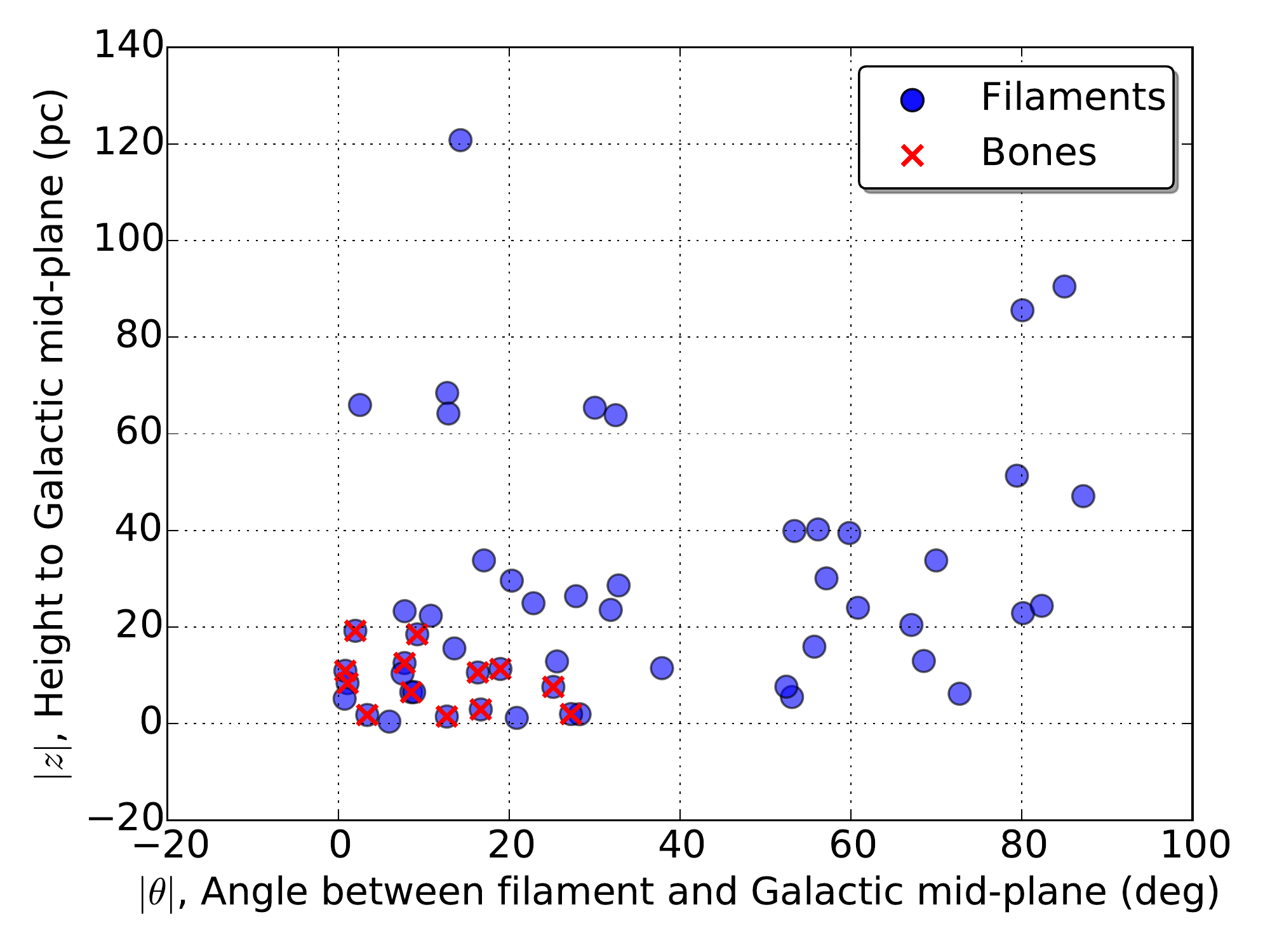}
\caption{
Filament orientation angle $|\theta|$ plotted versus the vertical height to the physical Galactic mid-plane $|z|$. The 54 filaments and 13 bones are labeled. For reference, the height of the Sun is $z_\odot = +25$ pc.
}
\label{fig:theta}
\end{figure}

Are filaments closer to the mid-plane more likely to align with the mid-plane, i.e., $\theta$ approaches $0^{\circ}$? The data does not indicate so (\autoref{fig:theta}):
the Pearson's product-moment correlation coefficient
between $|z|$ and $|\theta|$ is $C_{\rm Pearson} = +$0.26 or $-$0.35 for the filaments and bones respectively, far from a linear correlation ($+$1 or $-$1).

The statistical trends observed in these filaments provide excellent targets for quantitative tests with future theoretical calculations and numerical simulations.

\section{Discussion} \label{sec:discuss}

\subsection{Comparison to previously known filaments} \label{sec:knownFL}

Our MST algorithm finds some filaments previously identified by other methods.
Our search field (\autoref{sec:data}) partially overlaps with previous searches by \cite{Ragan2014-GFL}, \cite{me15}, and \cite{Zucker2015}.

Of the 9 cold and dense prominent filaments presented by \cite{me15}, 7 are in our search field, 3 of which are identified by MST (F7, F25, and F41 correspond to G11, G24, and G49, respectively).
Others are not identified because of a lack of dense BGPS sources (G29, G47, and G64) or too large disruption in velocity space due to active star formation (G28).

Among the 10 ``bone'' candidates presented by \cite{Zucker2015}, 6 are in our search field, 2 of which have dense BGPS sources: BC011.13--0.12 and BC024.95--0.17.
The former corresponds to F7 (the Snake), and the latter is not identified by MST because of too large velocity disruption.
In addition, our MST filament F28 is visible in their Figure 13 and seems to fulfill all their criteria, but was not identified by \cite{Zucker2015}.

Among the 7 giant molecular filaments presented by \cite{Ragan2014-GFL},
6 are partially covered in our field (``partially'' because most of those filaments extend beyond our coverage in $|b|$).
F36 is a small dense part of GMF38.1-32.4a, but note that \cite{Ragan2014-GFL} used a kinematic distance of 3.3--3.7 kpc, a factor of 2 larger than the ML distance of 1.7 kpc.
F19 and F38 fall in the positional coverage of GMF20.0-17.9 and GMF41.0-41.3, respectively,
but outside the velocity ranges.

In addition, F33 is the dense part of the ``massive molecular filament'' G32.02+0.06 presented by \cite{Battersby2012-FL} in a case study.
F13 is part of the IRDC G14.225-0.506 \citep{Busquet2013}.

F31 runs across a well studied IRDC \object{\ga}, also known as the ``Dragon'' nebula \citep{me11,me12,my-Springer-sum}. The IRDC is the IR-dark and submm-bright arc, bent towards the bottom of the panel in \autoref{fig:rgb}. The MST filament F31 is a new filament that runs across the IRDC at P1, where a proto-cluster is forming \citep{me11,me12,qz15}. Interestingly, the clump scale magnetic fields \citep{me12} are aligned with F31. At scales of the order of 10 pc, magnetic fields may be shaped by gravity, while on smaller scales (within 1 pc, or clump-scale), the magnetic fields control the formation of a secondary filament, as interpreted in \cite{me12}. The secondary filament is a small part of F31. Dust polarization observations of these filaments are needed to further investigate the role of magnetic fields on the formation and evolution of these filaments.

In summary, our MST method successfully finds previously known filaments where the criteria are satisfied. An important difference between the MST identified filaments and others is that the former contains dense clumps over the \textit{whole} extent, while this is not the case for previously ``by-eye'' identified large filaments (``gaps'' in velocity space are allowed).
The MST method also finds filaments embedded in a crowded PPV space, which are difficult to isolate with human eyes (e.g., F31).

It is noteworthy that, because of the filamenatry and hierarchical nature of the ISM, one can find an arbitrary number of filaments in the same data set using different criteria. Therefore, when presenting a filament sample, it is \textit{equally important} to explicitly list the criteria used to define filaments. For the same reason, when comparing different samples of filaments one has to notice the difference in criteria, otherwise the comparison is misleading.

\subsection{Completeness and bias} \label{sec:bias}

The 54 filaments form the first comprehensive sample of large-scale velocity-coherent gas structures in the northern Galactic plane covered by the BGPS spectroscopic survey. The homogeneous sample allows us to investigate statistical trends (\autoref{sec:stat}) for the first time. 
With length in the range of 10--276 pc and average column density above $10^{21}$ \cms, these filaments are among the densest and largest structures observed in the Galaxy, and provide excellent tracers for Galactic structure and kinematics \citep[e.g.][]{Englmaier1999,Dame2001,Dobbs2012,Reid2014,Vallee2016,Smith2016}.

The completeness of our filament sample largely depends on the data we use. The BGPS continuum catalog is 98\% complete at the 0.4 Jy level \citep{Rosolowsky2010_BGPS2}. The spectroscopic catalog \citep{Shirley2013} contains 50\% of the sources in the survey coverage with dense gas lines detected. As our identification used the spectroscopic catalog, the results are biased to dense clumps. This is evident in the high averaged column density and high linear mass density.
However, we emphasize that this is indeed our goal --- we are interested in the most prominent dense filaments.

On the other hand, given the location of our Sun in the Galaxy, even homogeneous surveys like BGPS or ATLASGAL are biased to structures (a) closer to the Sun and (b) on the same side with respect to the mid-plane as the Sun (\autoref{fig:hist}; \citealt{ATLASGAL}, but see discussion in \citealt{Rosolowsky2010_BGPS2}).
As mentioned in \autoref{sec:stat}, the large filaments show less bias than BGPS sources in $z$, but the distribution of $z$ is indeed not perfectly symmetric. Should we mirror the distribution of $z>0$ to $z<0$, the total number of filaments would increase to 74. That is, 27\% of the filaments may be missed due to this effect.

Although with a much improved completeness compared to previous methods,
the MST approach cannot find all large filaments in our Galaxy. On of the main strength of this method is the repeat-ability compared to manual approaches.

\begin{figure}
\centering
\includegraphics[width=.45\textwidth,angle=0]{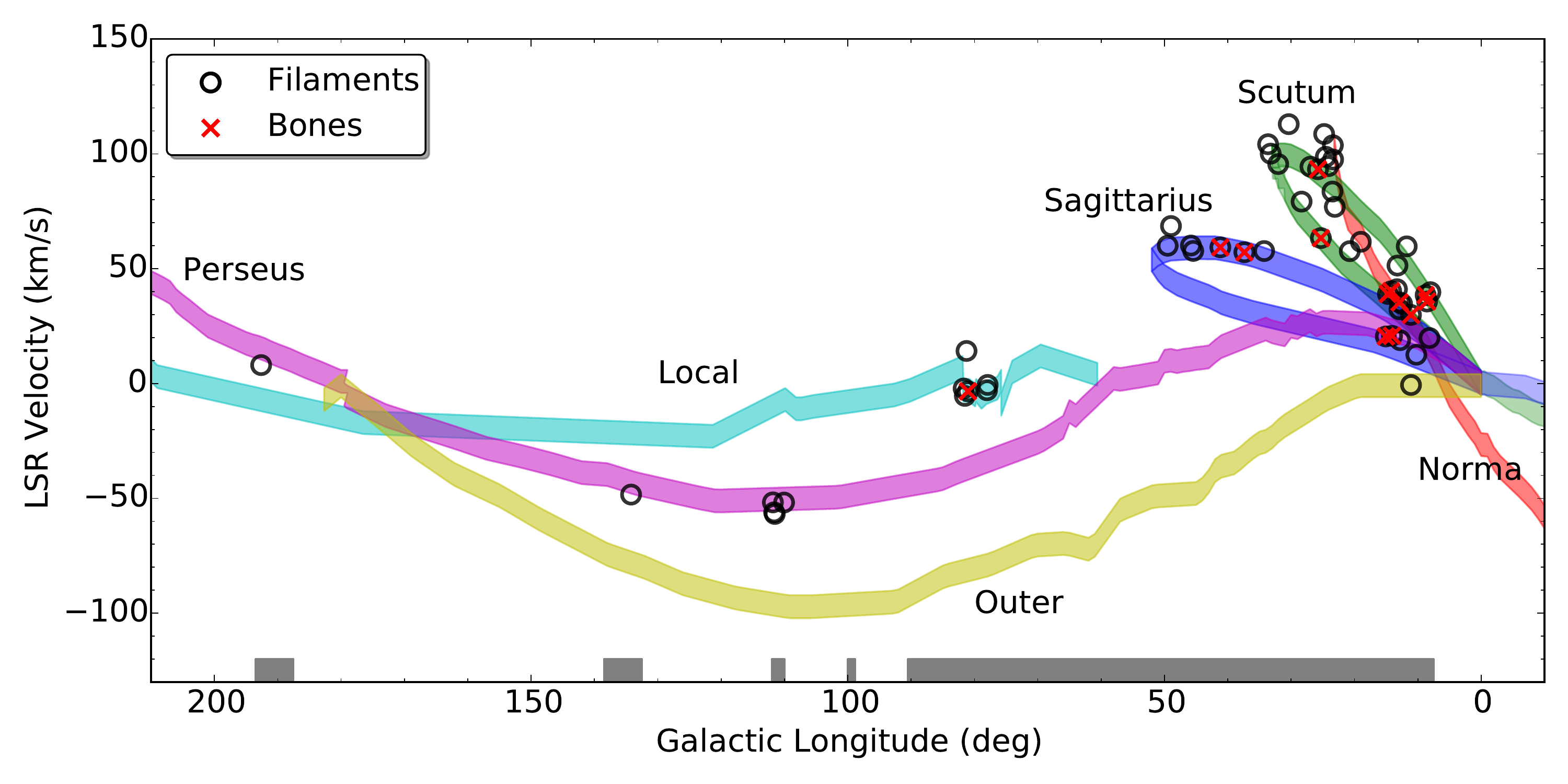}
\includegraphics[width=.45\textwidth,angle=0]{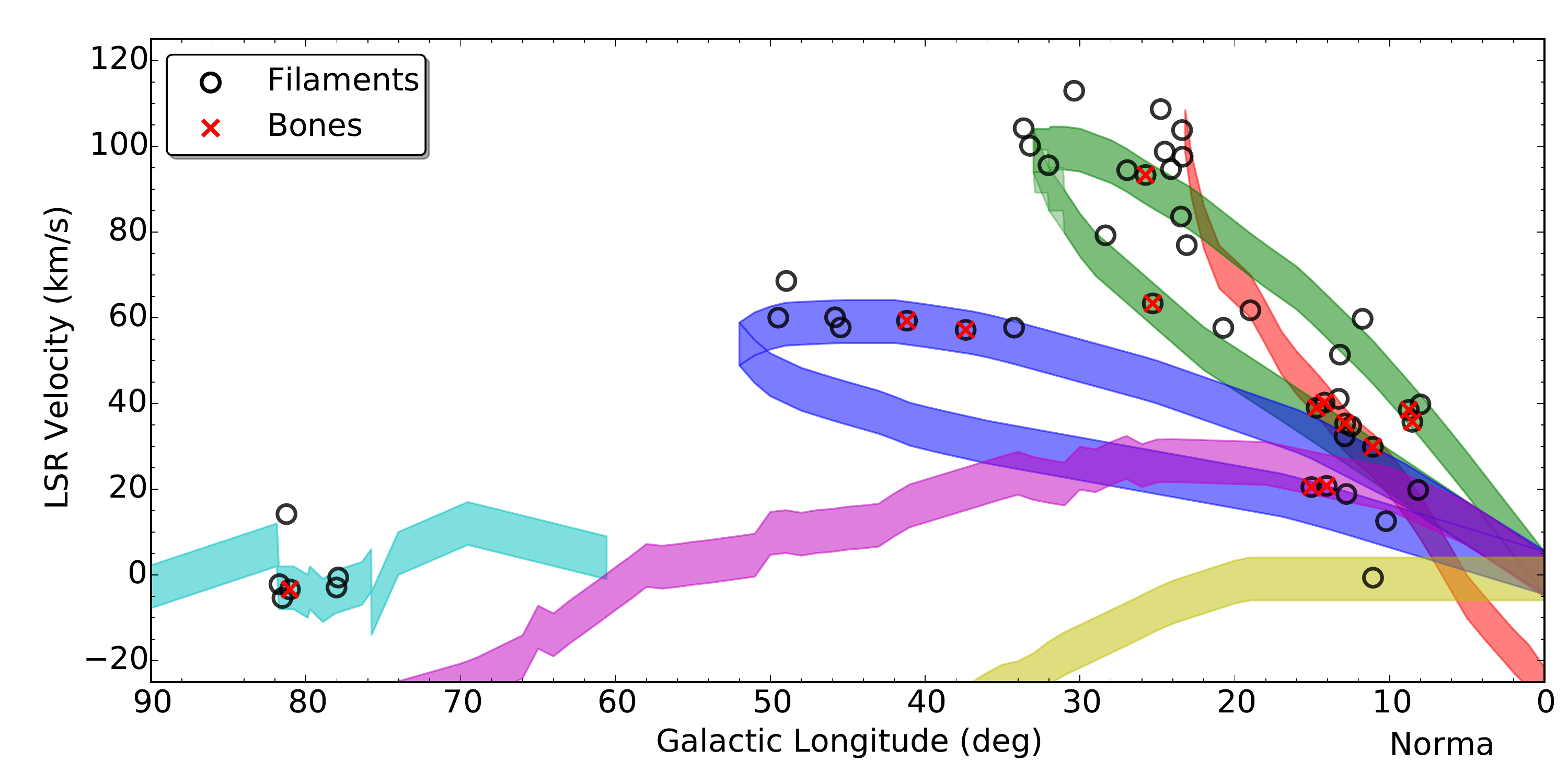}\\
\includegraphics[width=.43\textwidth,angle=0]{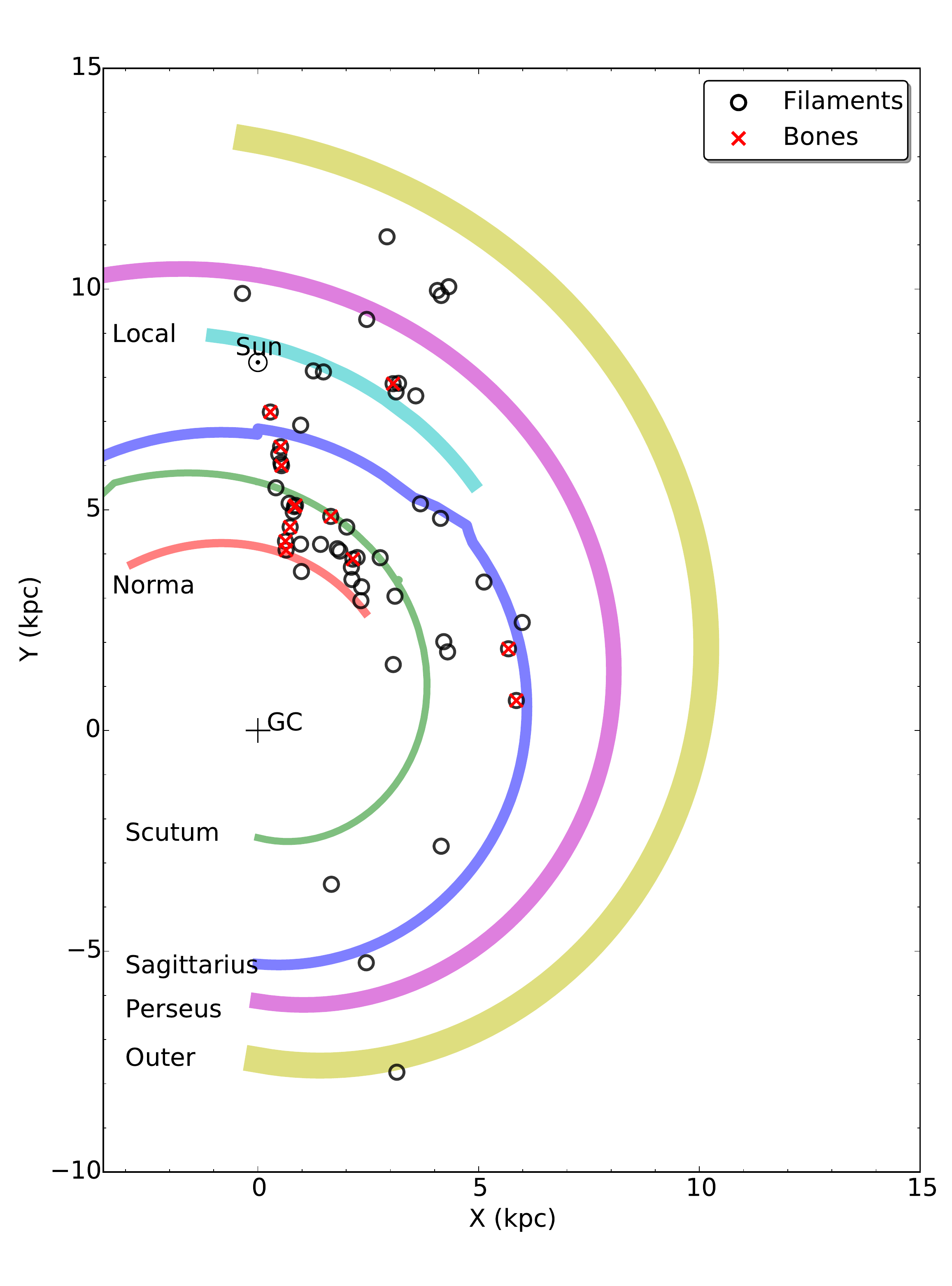}
\caption{
The Galactic distribution of the filaments.
\textit{Upper:}
The longitude-velocity plot showing the spiral arm segments derived from maser parallaxes (\citealt{Reid2014}; Reid et al. 2015, private communication). For simplicity, only related arms (Scutum, Sagittarius, Norma, Local, Perseus, and Outer) are plotted. The color shaded segments are of $\pm 5$ \kms\ width with respect to arm centers. The 54 filaments and 13 bones are labeled. The grey shaded horizontal strips along the x axis depict the searched longitude ranges. 
\textit{Middle:} a zoom-in of the upper panel for clarity.
\textit{Lower:}
A ``face-on'' view from the northern Galactic pole. The arm widths (170--630 pc) are from \cite{Reid2014} except for Norma, whose width is not available, and we plot 200 pc width for reference. The solar symbol $\odot$ is plotted at (0, 8.34) kpc.
}
\label{fig:arm}
\end{figure}

\subsection{Galactic distribution and number of large filaments and ``bones'' in the Galaxy}
\label{sec:Gal-dist-bones}

Most of the filaments are associated with major spiral arms (\autoref{fig:arm}), consistent with the observations by \cite{me15}.
Many of them concentrate along the longitude-velocity tracks of the Scutum, Sagittarius and Norma arms, and a few associated with the Local arm, the Perseus arm, and one associated with the Outer arm\footnote{In the $l-v$ view (\autoref{fig:arm}), most filaments follow the spiral arms, while the association is less evident in the face-on view. This originates from the difference in the the distance determination methods for the arm segments (parallax measurements) and the filaments (mainly kinematic distance). The same is seen in e.g. \cite{Abreu2016}.}.
Only a small fraction (11/54, or 20\%)
of the filaments are not within $\pm 5$ \kms\ of any arm structure, and are analogs of ``spurs'' observed in other galaxies.

How many large filaments exist in our Galaxy? Using our method, we have identified 48 filaments in the contiguous coverage of $7.5^{\circ} \le l \le 90.5^{\circ}$. It is reasonable to estimate a similar number of filaments in the fourth quadrant. In the outer Galaxy, the BGPS survey is targeted to several star formation regions, therefore the 6 identified filaments provide an extremely lower limit. 
Taking all these into account, and correcting for the bias as discussed in \autoref{sec:bias}, we estimate about 200 velocity-coherent filaments longer than 10 pc and with a global column density above $10^{21}$ \cms, as the filaments presented in this study.

How many filaments lie in the center of spiral arms and thus sketch out the ``bones'' of the Milky Way? \cite{Goodman2014} argued that the long and skinny IRDC ``Nessie'' lies in the center of the Scutum-Centaurus spiral arm in the $(l,v)$ space, and within $z = \pm 20$ pc from the physical mid-plane.
Following \cite{Goodman2014} and \cite{Zucker2015}, our criteria for a ``bone'', on top of our large-filaments criteria (1--5) (\autoref{sec:method}), are:
\begin{enumerate}
\item[(6)]
Lies in the very center of the physical Galactic mid-plane, with
$|z| \le 20$ pc. 
\item[(7)]
Runs almost parallel to arms in the projected sky, with 
$|\theta| \le 30^{\circ}$.
\item[(8)]
The flux weighted LSR velocity $v_{\rm wt}$ is within $\pm 5$ \kms\ from spiral arms.
\end{enumerate}
However, the exact structure and position of the spiral arms in our Galaxy are not well established. Diverse models have been derived from a variety of data ranging from atomic, molecular, ionized gas to stars and pulsars \citep[e.g.][]{Reid2014-AR,Hou2014-arm,Vallee2015,Vallee2016}.
{Here we}
have adopted the spiral segments derived from maser parallaxes (\citealt{Reid2014}; Reid et al. 2015, private communication), which have well constrained distances.
In \autoref{fig:arm} we superpose the filaments on the spiral segments.
Among the 54 filaments, 27 fulfill criteria (1-6), 21 fulfill criteria (1-7), and 13 of them also fulfill criterion (8). These 13 filaments (F2, F3, F7, F10, F13, F14, F15, F18, F28, F29, F37, F38, and F48) are ``bones'' according to our definition.
Our criteria for a bone are more strict than \cite{Zucker2015} in terms of velocity coherence and mean column density. 
When compared to other filaments, bones do not stand out in mass, length  (\autoref{fig:hist}(a--b)), column/volumn density, or temperature.
All the 13 bones are located in the first quadrant (which is not surprising given our search field, see \autoref{sec:data}), making 27\% of the 48 filaments in the same region of blind BGPS survey.
Obviously, owning to disagreement among the many spiral arm models, adopting a different model will lead to different ``bones''. But the fraction of bones in filaments should not change dramatically in a reasonable spiral model.

\subsection{Fraction of ISM confined in large filaments} \label{sec:fractionISM}

Given the importance of filamentary geometry in enhancing massive clustered star formation,
it is of great interest to quantify the fraction of the ISM contained in large filaments and to evaluate the star formation activities therein.

To address this question we consider only the range of $7.5^{\circ} \le l \le 90.5^{\circ}$ where the BGPS and its spectroscopic follow-up are contiguous.
In this field there are 5841 BGPS v1 sources, 2893 having \hcop or \NTH (3--2) detection which we term as ``dense BGPS sources''.
We identified 48 filaments in this field, which are comprised of 521 BGPS sources. That means 17.7\% (512/2893) of dense BGPS sources, or 8.8\% (512/5841) of all BGPS sources, are confined in large filaments.
If we count BGPS clumps in the bones only, 6.8\% of dense BGPS sources, or 3.4\% of all BGPS sources are confined in bones.

Compact 1.1 mm continuum emission of the BGPS sources outline the dense, inner part of much larger and less dense envelopes of molecular clouds \citep{Dunham2011-BGPS-NH3}. Assuming a dense gas mass fraction of 10-20\% \citep[cf.][]{Ragan2014-GFL,Ginsburg2015-W51}, we infer an order of 1\% of the molecular ISM is confined in large filaments, and about 1/3 of this amount confined in bones -- marking spiral arm centers.

\begin{figure*}
\centering
\includegraphics[width=.46\textwidth,angle=0]{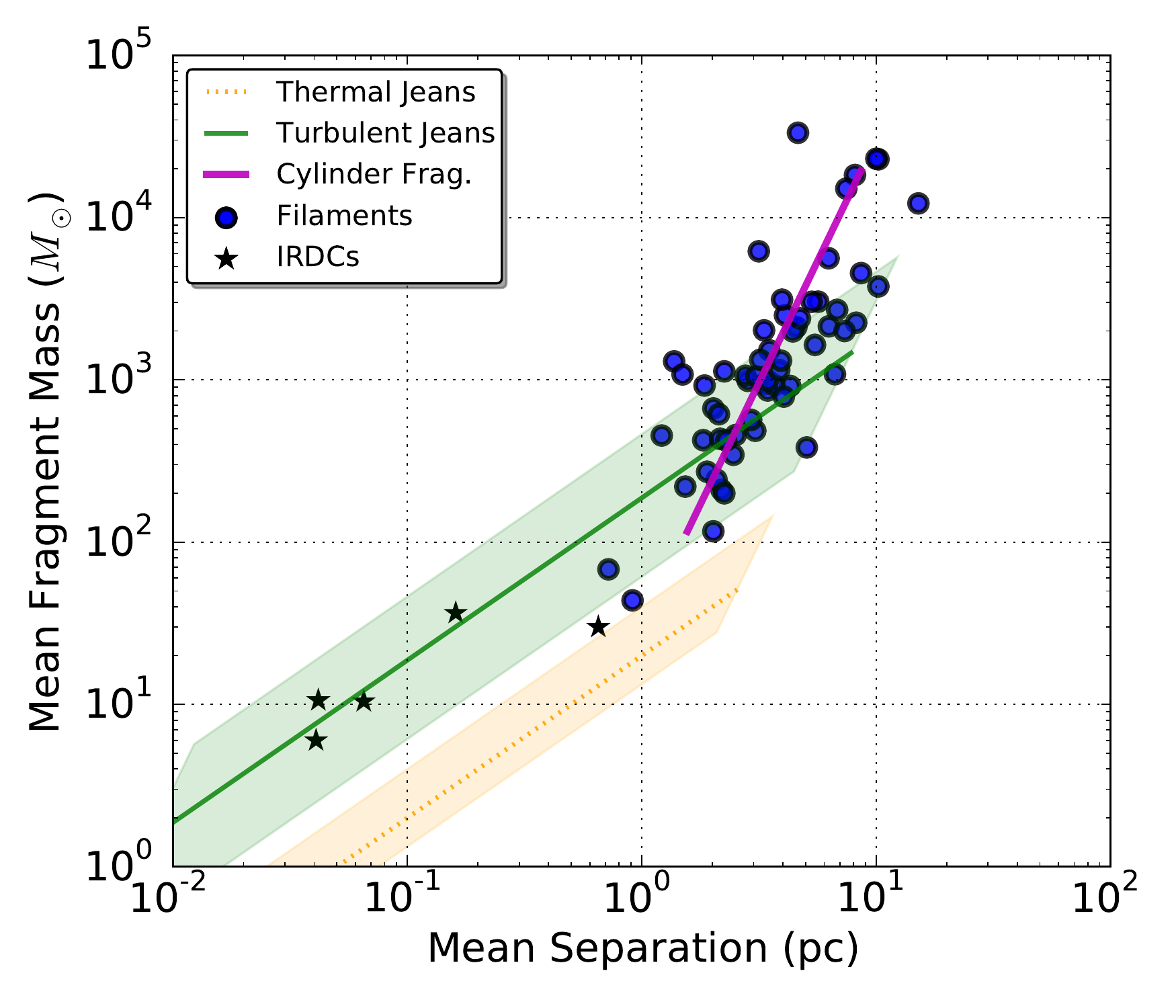}
\includegraphics[width=.52\textwidth,angle=0]{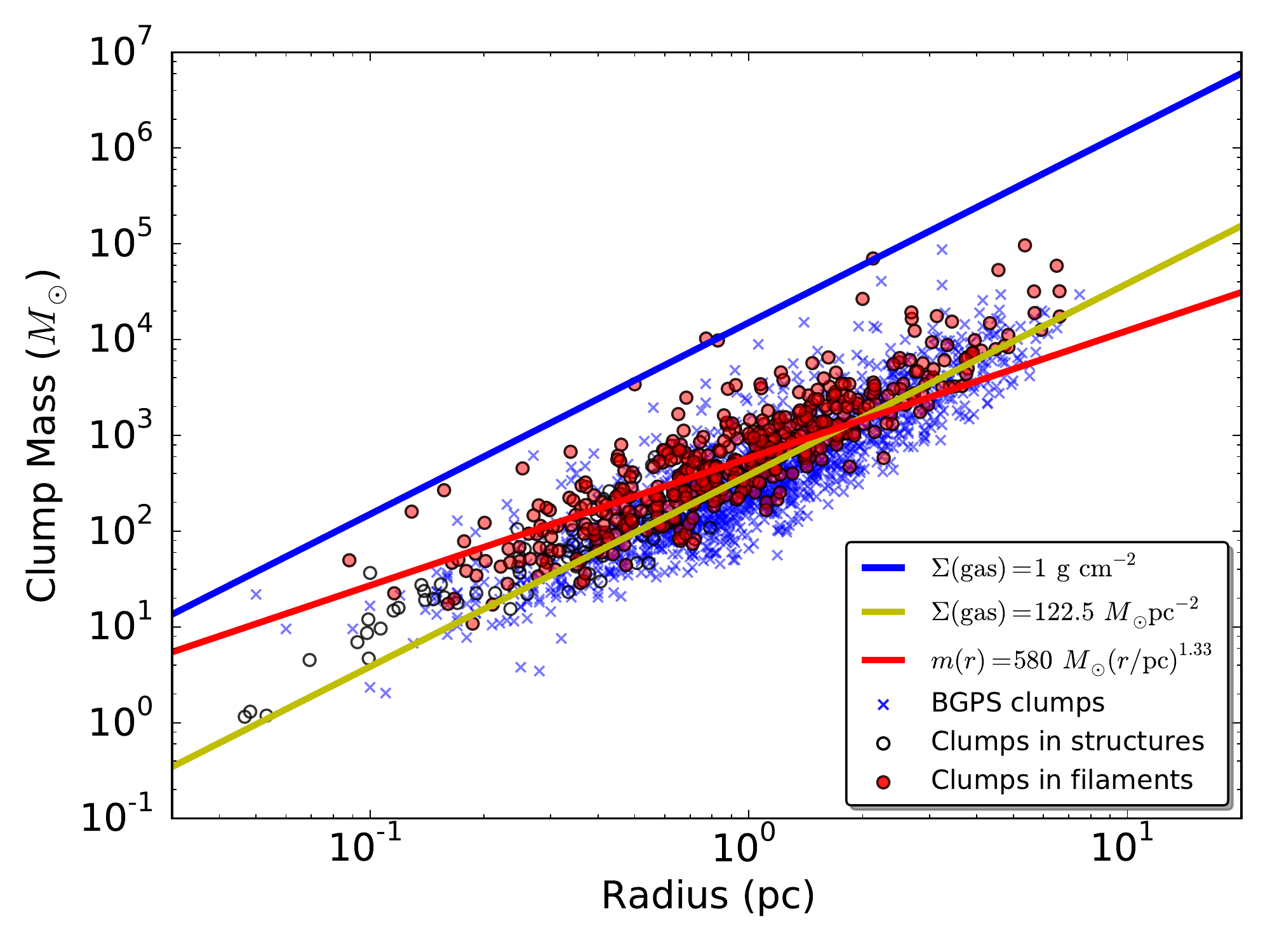}
\caption{
\textit{Left:}
Mean clump mass versus mean edge length for the 54 filaments, as compared to IRDCs and theoretical predictions.
The magenta line  is \textit{not} a fit to the data points, instead, it depicts 
{cylindrical}
fragmentation appropriate for the filaments (see text). In comparison, the resolved fragmentation of IRDCs \citep[data from][]{me11,me12,me14,qz11} is consistent with turbulent Jeans fragmentation. The green line, orange line, and associated shaded regions correspond to a range of density and temperature appropriate for IRDCs \citep{me14}.
\textit{Right:}
Clump mass versus radius for BGPS clumps, clumps in large filaments, and clumps in other velocity-coherent structures. The color lines depict various empirical criteria of star formation: blue -- the \cite{Krumholz2008Nature} criterion of massive star formation;
yellow -- average of \cite{Heiderman2010} and \cite{Lada2010}  for ``efficient'' star formation;
red -- the \cite{Kauffmann2010} criterion of massive star formation with a correction of the
{adopted}
dust opacity law as in \cite{Dunham2011-BGPS-NH3}.
}
\label{fig:frag}
\end{figure*}

\subsection{An apparent length limit of 100 pc and the longest filaments beyond this limit}
\label{sec:100pc}

\cite{me15} pointed out an apparent upper limit of 100 pc projected length for the longest filaments in their study designed to find cold and dense filaments based on \her\ far-IR emission.
In this study, we use a different approach without limiting the temperature. Except the extremely long filament F5, all other filaments are indeed shorter than 100 pc. Interestingly, this limit is also seen in \cite{Zucker2015} despite of the different search method. The 100 pc limit seems to be present in filaments with a global column density above the order of $10^{21} - 10^{22}$ \cms. Relaxing the column density to a lower cut of $<10^{20}$ \cms, longer filaments start to be picked up in $^{12}$CO/$^{13}$CO (1--0): the the 430 pc ``optimistic Nessie'' \citep{Goodman2014}; the 500 pc ``wisp'' \citep{LiGX2013-FL500pc}, and few filaments by \citealt{Ragan2014-GFL} and \citealt{Abreu2016}. However, those CO filaments have much smaller aspect ratios (typically $\ll$10, see figures in their papers), and the low-$J$ CO gas outlines the relatively diffuse envelopes of denser structures traced by MIR extinction. For example, one filament reported by \cite{Ragan2014-GFL} contains F36 as a small part (\autoref{sec:knownFL}).
Whether 100 pc is a true limit for dense filaments warrants further study. This provides a quantitative test case for numerical simulations \citep[e.g.][]{Falceta2015}.

So far, the 276 pc long filament F5 (\autoref{fig:rgb}) is the only exception of dense ($>10^{21}$ \cms) filaments longer than 100 pc. Compared to the above mentioned extremely long CO filaments, F5 is at least 10 times denser in average column density, and it may also have less dense extensions similar to the 80 pc ``classic Nessie''.
More systematic searches and comparison to numerical simulations can resolve the true length limit of the longest filaments.
We {emphasize}
that average column density is a crucial parameter in defining the boundary of filaments
and thus the length and aspect ratio. It is also worthwhile to note that our filaments, as defined by a collection of dense BGPS clumps, form the center of larger and less dense structures.

The origin of large velocity coherent filaments is still a mystery. Numerical simulations
of the multiphase ISM in galactic disks demonstrate that the cold, dense gas component tends
to organize itself naturally into a filamentary network 
\citep[e.g.][]{Tasker2009,Smith2014}.
Spiral arms can sweep up and compress gas, generating bones \citep{Goodman2014}. Gravitationally unstable disk regions condense into gaseous rings and arcs \citep{Behrendt2015}. In differentially rotating disks, structures like molecular cloud complexes could be sheared into elongated filaments.

Our calculations (Burkert et al. in prep) show that tidal effects of the Milky Way are too weak to affect the maximum length of filaments. This is consistent with our observations, where we find filament lengths do not correlate with Galactocentric radii ($C_{\rm Pearson} = -0.07$).
The maximum filament length of order 100 pc might be related to the timescale of $\tau_{\rm SF} = 10^7$ yrs \citep[e.g.][]{Burkert2013} on which stars form in dense molecular gas and destroy their environment. Typical turbulent velocities on large scales in galactic
disks are of order $\sigma = 10$ \kms\ \citep{Dib2006}. If $\sigma$ is the maximum velocity
with which coherent filaments can grow and if $\tau_{\rm SF}$ denotes the timescale on which they are destroyed again their length is limited by $l = \sigma \times \tau_{\rm SF}$ = 100 pc, in agreement with the observations.

\subsection{Fragmentation of large-scale filaments and subsequent star formation} \label{sec:frag}

By definition, the filaments presented in this study are in the form of a chain of dense clumps physically connected by less dense gas in between. For linear filaments, this geometry resembles a fragmented ``cylinder'' with regularly spaced clumps under the ``sausage instability'' of self-gravity (e.g. F9 in \autoref{fig:rgb}). According to the framework of \cite{Chandra1953}, an isothermal gas cylinder becomes super-critical when its linear mass density exceeds the critical value $(M/L)_{\rm crit}$, and will fragment into a chain of equally spaced fragments with a spacing of $\lambda_\mathrm{cl}$, with each fragment having a mass of $M_\mathrm{cl} = (M/L)_{\rm crit} \times \lambda_\mathrm{cl}$. 
In short, the fragmentation is governed by central density and pressure (thermal plus non-thermal).
This framework has been followed by many authors \citep[e.g.][]{Ostriker1964_FL,Nagasawa1987_FL,Bastien1991_FL,Inutsuka1992_FL,Fischera2012_FL}. See \cite{me11,me14} for useful deduction of the formulas. \autoref{fig:frag} (left panel) plots the mean clump mass of each filament with the mean separation between clumps (mean length of the edges in the filament). The observed fragmentation is consistent with the theoretical prediction of cylindrical fragmentation assuming a central density of $1\times 10^4$ \cmc\ and velocity dispersion of 0.4--2.2\kms\ (magenta line). 
The spread of the data points around the prediction line may be due to a range of central densities and imperfect cylinder geometry. Recent numerical simulations have shown that geometric bending, which is often seen in observed filaments, can change the regularity of the spacing \citep{Gritschneder2016}, indicating that more theoretical work is required in order to understand the stability and dynamics of filaments.

Dense clumps with a typical mass of $10^3$ \msun\ and typical size of 1 pc (\autoref{fig:frag}, left panel) are, in general, capable of forming a cluster of stars.
Statistically, dense clumps residing within large filaments are slightly denser than clumps elsewhere (see below). in \autoref{fig:frag} (right panel), we plot clump mass versus deconvolved radius (not all BGPS clumps have a valid radius, see \autoref{sec:para}) for three categories of BGPS clumps: I -- the 1710 clumps with well determined distance from \cite{Ellsworth2015}. II -- the 294 clumps in velocity-coherent structures but not large filaments. III -- the 469 clumps in large filaments. Categories I, II, and III have 41.0\%, 39.8\%, and 46.2\% clumps satisfying the \cite{Kauffmann2010} threshold of forming massive stars. Thus, categories I and II are indistinguishable, while in comparison, category III is slightly more favorable for massive star formation.
If we count in mass instead of number of clumps, the fractions are 79.2\% for BGPS sources, 86.3\% for velocity-coherent structures but not large filaments, and 91.0\% for large filaments.
Surprisingly, bones do not show a higher fraction compared to large filaments, either counted in number or mass.
This indicates that local environment such as a velocity-coherent filament
plays a role in enhancing massive star formation. Filaments, in particular, provide a preferred form of geometry to channel mass flows that can inhomogeneously feed star-forming clumps
\citep[e.g.][]{Arzoumanian2011,Peretto2014-SDC13,qz15,Heigl2016,Federrath2016}.
On the other hand, the Galactic environment does not seem to affect local star formation across the few hundreds pc spread of vertical position $z$, consistent with previous studies \citep{Eden2012,Eden2013}.

\section{Summary} \label{sec:sum}

We present an automated method designed to identify large-scale velocity-coherent filaments throughout the Galaxy. The method utilizes a customized MST algorithm, which connects neighboring voxels in the PPV space. We have applied the algorithm to the BGPS spectroscopic catalog in the range $7.^{\circ}5 \le l \le 194^{\circ}$, $|b|<0.5^\circ$.
We have identified a comprehensive catalog of 54 large-scale filaments and derived physical parameters including mass ($\sim 10^3 - 10^5$\msun), length (10--276 pc), linear mass density (54--8625 \msun\,pc$^{-1}$), aspect ratio (18--176), linearity, velocity gradient, temperature, fragmentation, Galactic location and orientation angle.
We investigate the Galactic distribution of these parameters and compare the filaments with an updated Galactic spiral arm model.

\begin{enumerate}
\item 
In general, the Galactic distribution of large filaments follows the dense gas traced by BGPS sources. However, there are interesting exceptions to be further explored by a quantitative comparison to theoretical work (\autoref{sec:stat}).

\item
Most of the filaments are associated with major spiral arms including the Scutum, Sagittarius and Norma arms, and a few associated with the Local, Perseus, and Outer arms. About 20\% of the filaments are inter-arm structures, or ``spurs'' (\autoref{sec:Gal-dist-bones}).
The filaments tend to align with Galactic plane, but the tendency does not correlate with vertical height (\autoref{fig:theta}).

\item 
The filaments are widely distributed across the Galactic disk, with 50\% located within $\pm$20 pc from the Galactic mid-plane and 27\% runs in the center of major spiral arms, forming the ``bones'' of the Milky Way (\autoref{sec:Gal-dist-bones}).

\item 
An order of 1\% of the molecular ISM is confined in large filaments, and about 1/3 of this amount is confined in bones -- marking spiral arm centers (\autoref{sec:fractionISM}).

\item 
Massive star formation is more favorable in large filaments compared to elsewhere.
However, Galactic environment is not observed to affect local star formation (\autoref{sec:frag}).

\item 
An apparent length limit of 100 pc is observed for filaments with a global column density $N_{\rm H_2}$ higher than $10^{21}$\cms\ (or optical extinction $A_V \approx 1$ mag). Longer filaments are rarer, with a much lower aspect ratio, and have at least one order of magnitude lower global column density (\autoref{sec:100pc}).

\end{enumerate}

Our method can be applied to 3-dimensional PPV catalogs from observations or PPP catalogs from simulations. This study focuses on the northern Galactic plane covered by the BGPS. In the near future, the 2-dimensional ATLASGAL (inner Galactic plane; \citealt{cat:ATLASGAL-Csengeri2014}) and Hi-GAL (full Galactic plane; \citealt{survey:HiGAL-DR1-Molinari2016}) catalogs will be complemented with \vlsr\ information from currently ongoing spectral line surveys
(SEDIGISM: Structure, Excitation, and Dynamics of the Inner Galactic Interstellar Medium, Schuller et al. submitted;
MWISP: Milky Way Imaging Scroll Painting, \citealt{Jiang2013-MWISP,Sun2015-arm};
Mopra Southern Galactic Plane CO Survey, \citealt{Burton2013-MopraCO};
and
ThrUMMS: Three-mm Ultimate Mopra Milky Way Survey, \citealt{Barnes2015-ThrUMMS}).
By then, we can complete the census of large-scale, velocity filaments in the full Galactic plane.

As increasing number of filaments with various morphologies are published in the literature, we urge our colleagues to explicitly list the criteria used to define their filaments before a commonly agreed, physics driven, definition of filaments can be reached in the community. 
The final definition will likely be scale dependent, given the different physics behind large-scale and smaller-scale filaments in the ISM.

\acknowledgments{
\textit{Acknowledgments}
We are grateful to Mark Reid and Tom Dame for providing the spiral arm segments, and Alberto Sanna for the parallax distance to maser G11.11-0.11, before publication.
We thank an anonymous referee for a constructive review report.
K.W. is supported by grant WA3628-1/1 of the German Research Foundation (DFG) through the priority program 1573 (``Physics of the Interstellar Medium'').
Color bars in the figures utilize the \textsc{cubehelix} color scheme introduced by \cite{Green2011-cubehelix}.
This research made use of \textsc{Astropy}, a community-developed core Python package for Astronomy \citep{Astropy}.
This publication makes use of data products from the Wide-field Infrared Survey Explorer, which is a joint project of the University of California, Los Angeles, and the Jet Propulsion Laboratory/California Institute of Technology, funded by the National Aeronautics and Space Administration.
}

\facility{CSO, SMT, \textit{Spitzer}, \textit{WISE}}
\software{\textsc{Python, Astropy}}

\begin{figure*}
\centering
\includegraphics[width=\textwidth]{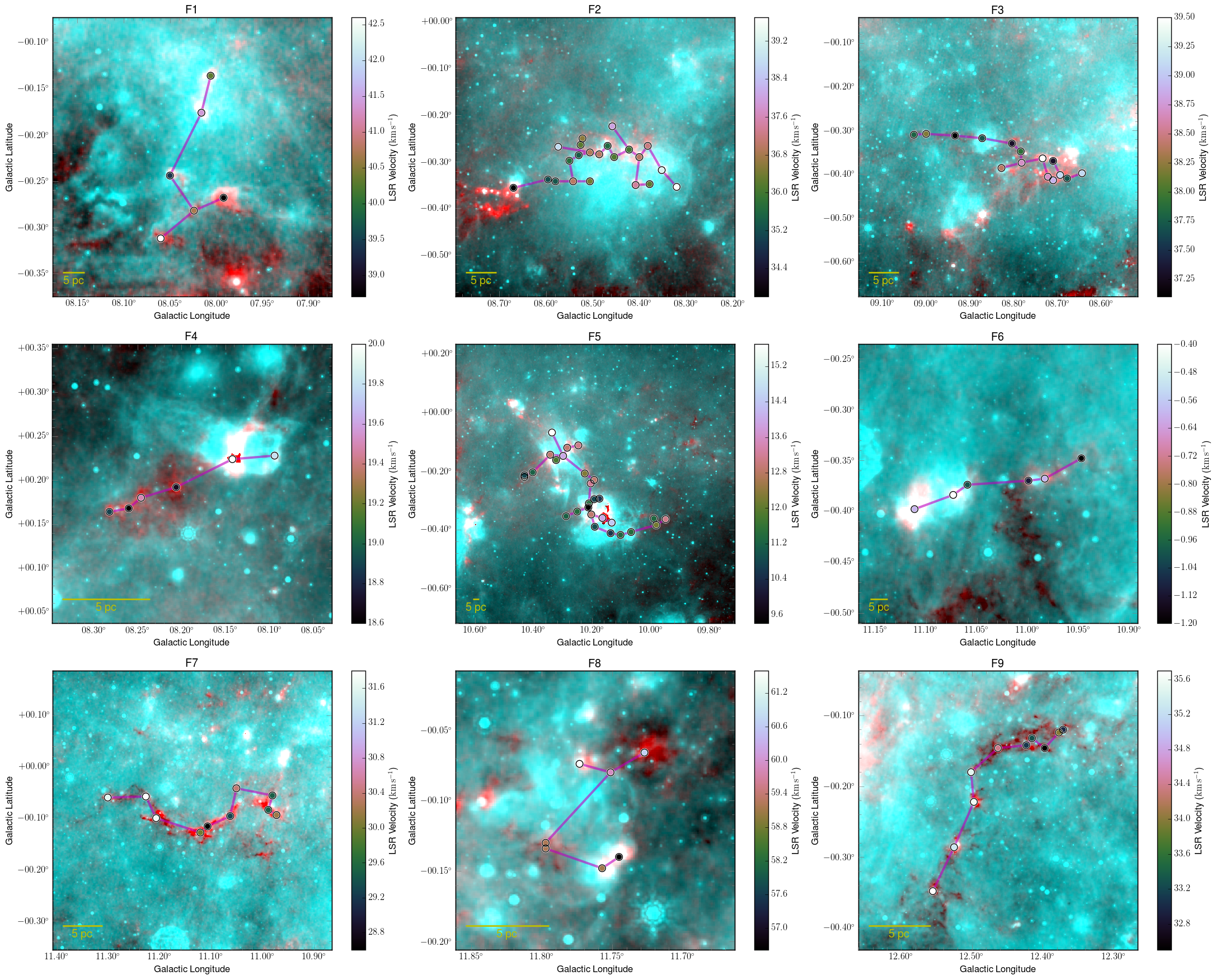}
\caption{A two-color view of the filaments. For F1 to F42, cyan shows the \spt\ 24\um\ emission in logarithmic scale \citep[from the MIPSGAL survey,][]{MIPSGAL}, and red shows the APEX+\textit{Planck} 0.87\,mm emission in linear scale \citep[from the ATLASGAL survey,][]{Csengeri2016-ATLASGAL}. For F43 to F54, cyan is \textit{WISE} 22\um\ emission \citep{WISE} and red is the BGPS 1.1\,mm emission \citep{Ginsburg2013_BGPSv2}.
The filled circles represent dense BGPS sources with color coded velocity as shown in the color bar. The sources are connected by MST edges (see text).
A scale bar of 5 pc is shown for reference.
}
\label{fig:rgb}
\end{figure*}

\setcounter{figure}{5}
\begin{figure*}
\centering
\includegraphics[width=\textwidth]{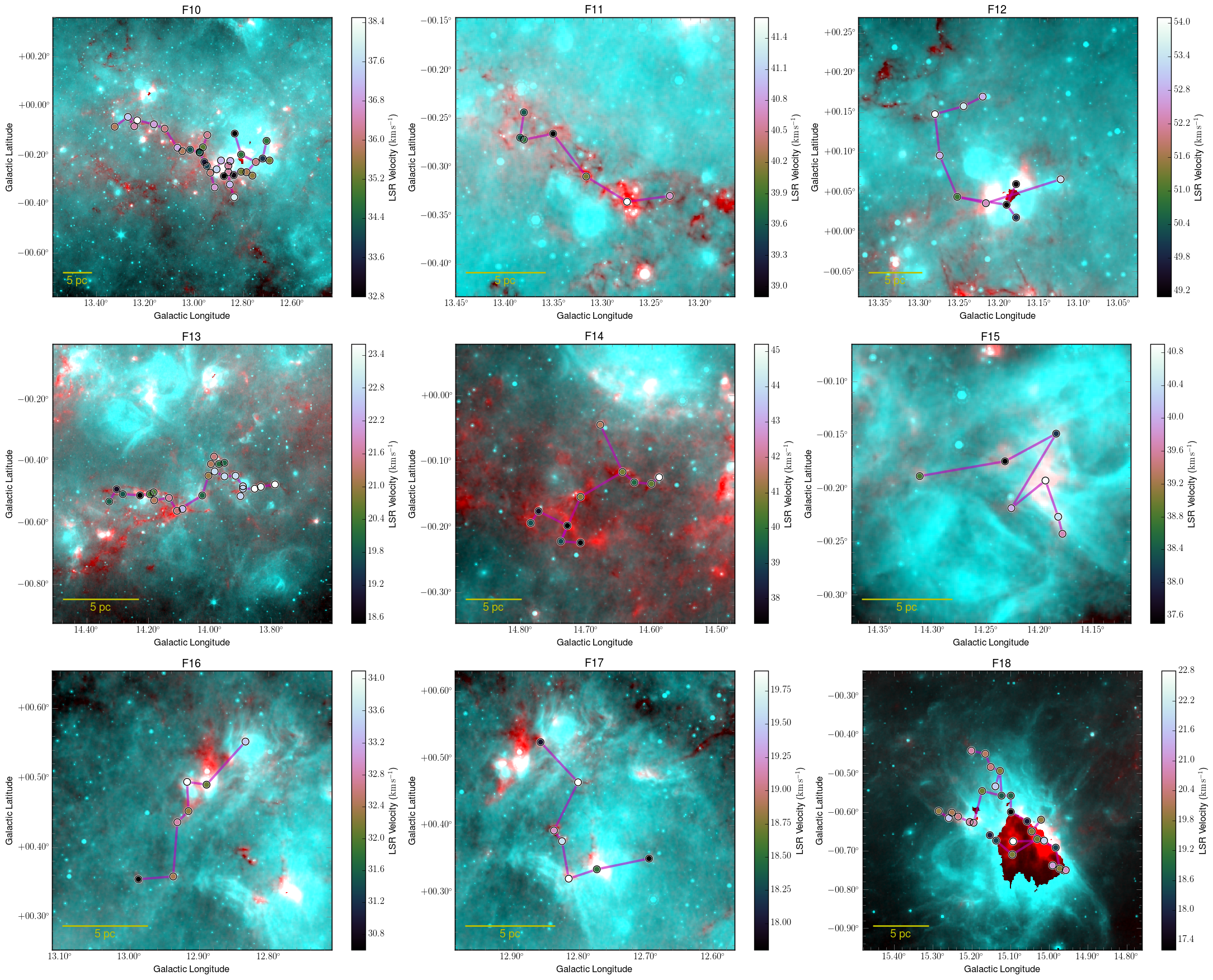}
\caption{Continued.
}
\end{figure*}

\setcounter{figure}{5}
\begin{figure*}
\centering
\includegraphics[width=\textwidth]{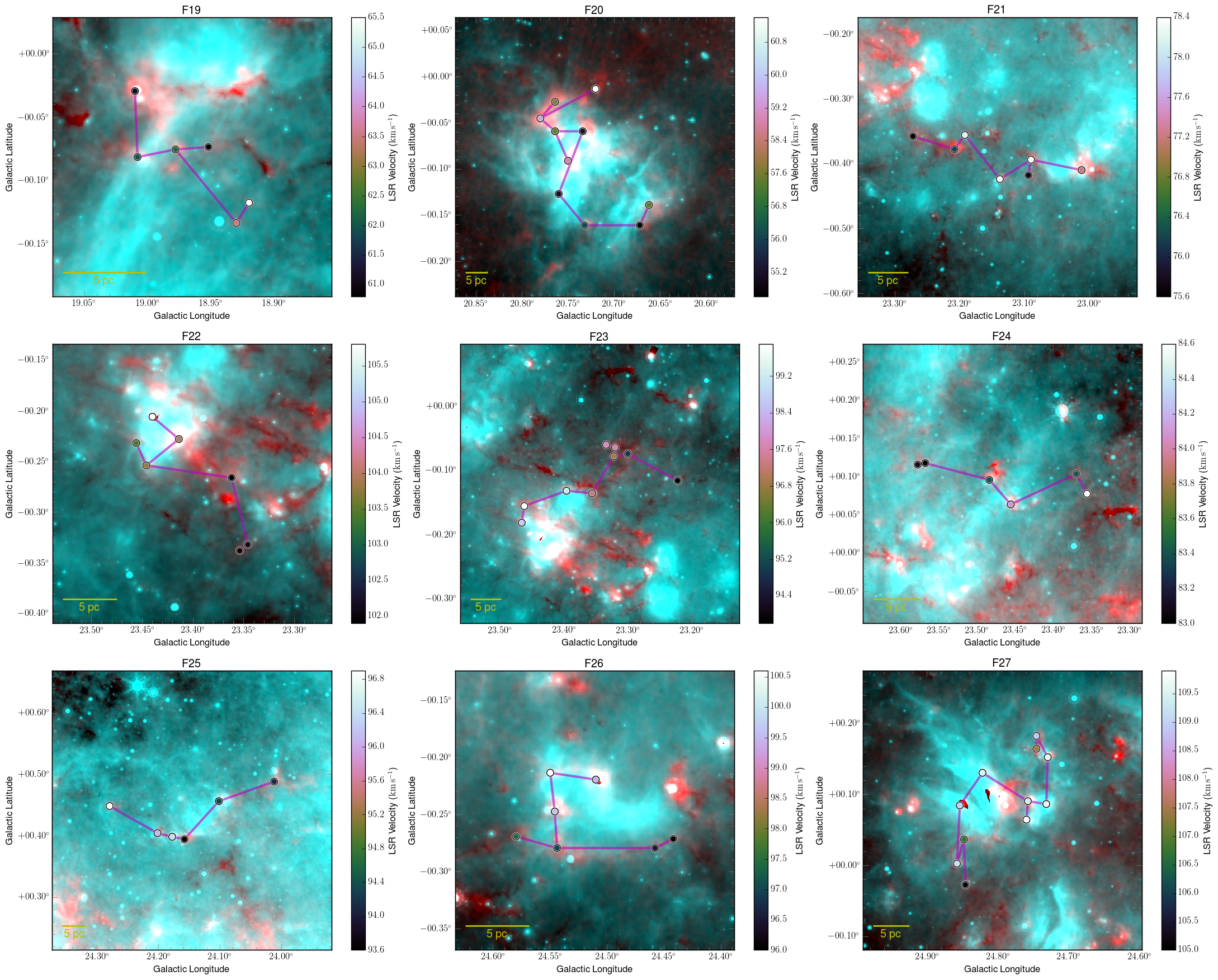}
\caption{Continued.
}
\end{figure*}

\setcounter{figure}{5}
\begin{figure*}
\centering
\includegraphics[width=\textwidth]{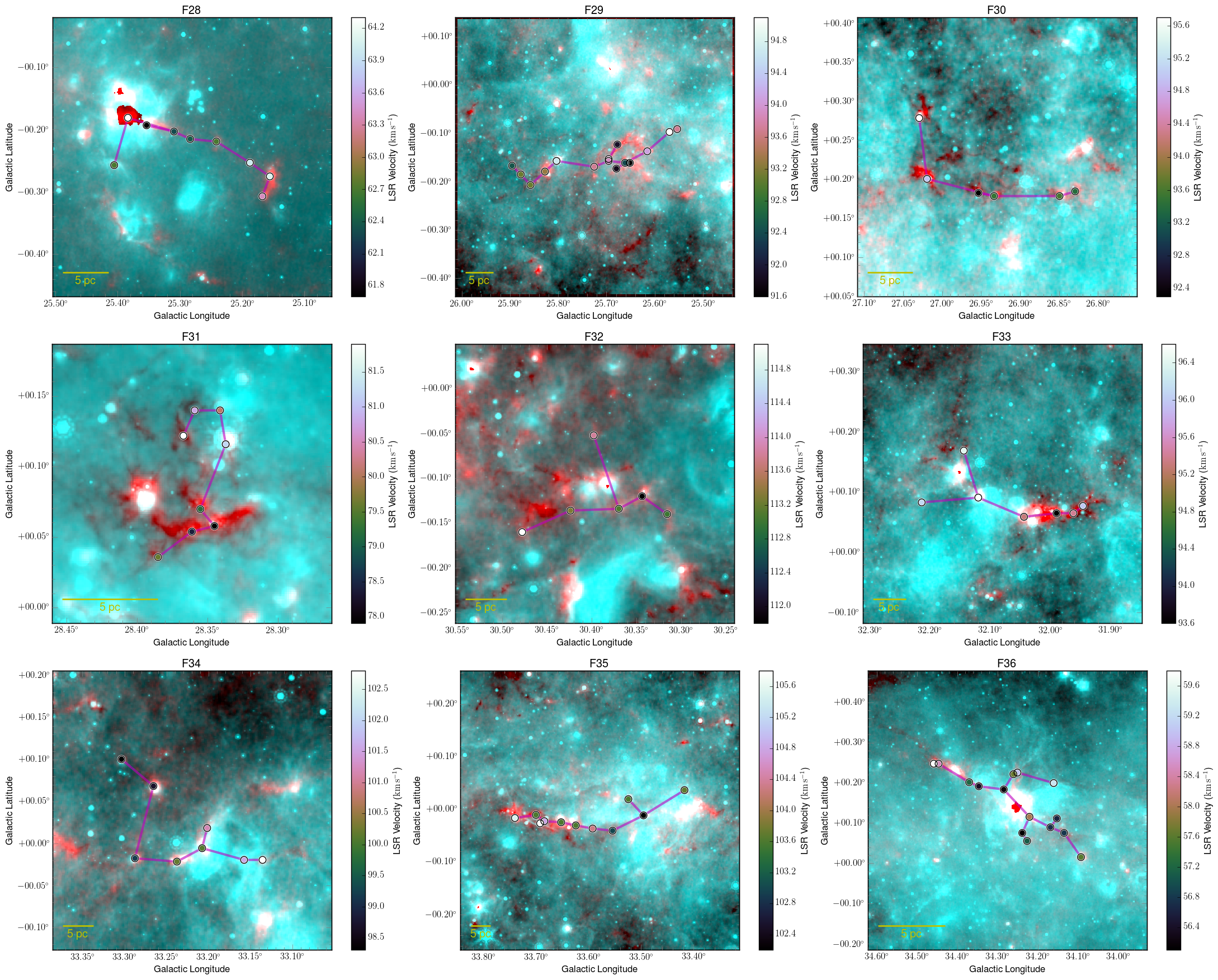}
\caption{Continued.
}
\end{figure*}

\setcounter{figure}{5}
\begin{figure*}
\centering
\includegraphics[width=\textwidth]{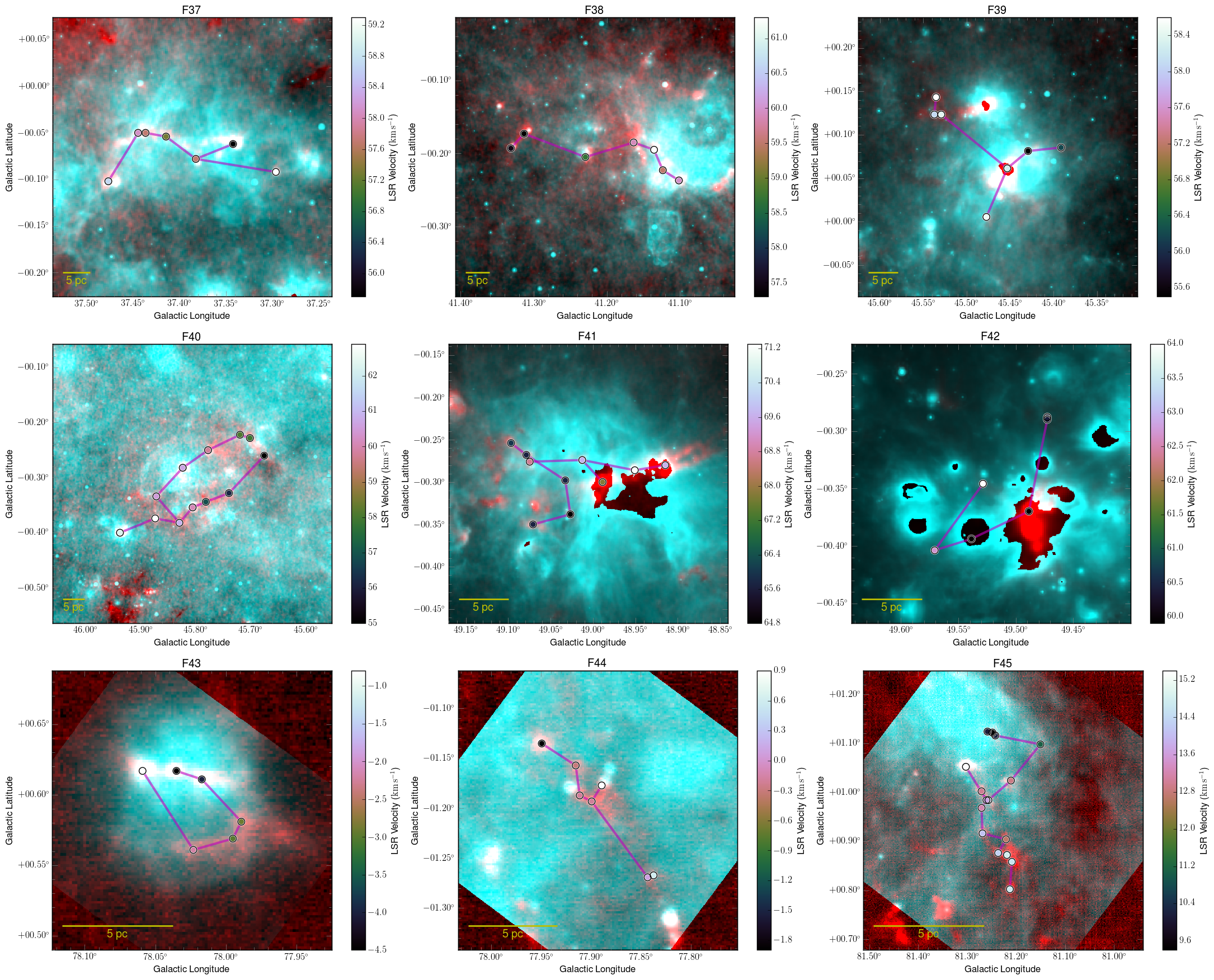}
\caption{Continued.
}
\end{figure*}

\setcounter{figure}{5}
\begin{figure*}
\centering
\includegraphics[width=\textwidth]{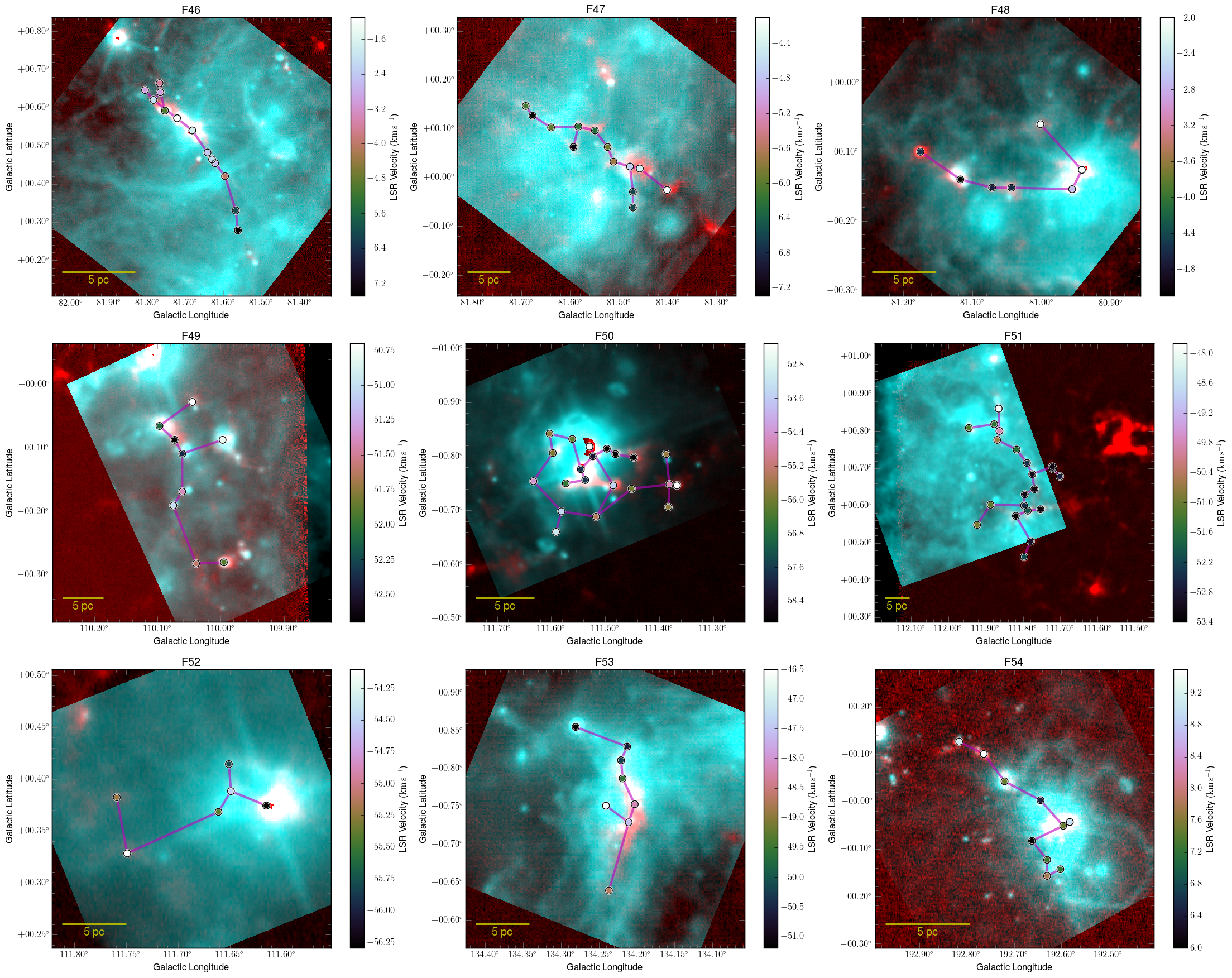}
\caption{Continued.
}
\end{figure*}


\setlength\tabcolsep{4.5pt}
\clearpage
\begin{turnpage}
\capstartfalse
\begin{table}
\caption{Physical parameters and statistics \label{tab:fl}}
\scriptsize
\begin{tabular}{lcccc cccc ccccc ccccc ccccc ccc}
\hline\\
{(1)} 
& {(2)} & {(3)} & {(4)} 
& {(5)} & {(6)} 
& {(7)} 
& {(8)} & {(9)} 
& {(10)} & {(11)} 
& {(12)} & {(13)} 
& {(14)} & {(15)} 
& {(16)} & {(17)}
& {(18)} & {(19)}
& {(20)} & {(21)} & {(22)}
& {(23)}
& {(24)} & {(25)} & {(26)} & {(27)}
\\
{ID} 
& {$l_{\rm wt}$} & {$b_{\rm wt}$} & {$v_{\rm wt}$} 
& {$d$} & {$d_{\rm type}$} 
& {$N_{\rm cl}$} 
& {$L_{\rm deg}$} & {$L_{\rm pc}$} 
& {$\overline{(\frac{\Delta v_i}{\Delta L_i})}$} & {$\sigma_v$} 
& {$T_{\rm min}$} & {$T_{\rm max}$} 
& {Mass} & {$M/L$}
& {$N_{\rm H_2}$} & {$n_{\rm H_2}$} 
& {$f_A$} & {$f_L$} 
& {$R_{\rm gc}$} & {$z$} & {$\theta$}
& {Morph.} 
& {$l_{\rm min}$} & {$l_{\rm max}$}
& {$b_{\rm min}$} & {$b_{\rm max}$}\\
\hline \\
\object{F1}   &   8.01 & -0.24  &  39.8 & 11.9 &   ML &    6 &  0.25 &   51.1 &  0.17 &  1.27 & 14.5 & 30.7 &  22.6 &  443.4 &  0.57 &  0.48  &   30.4 &   3.0 &  3.9 &  -51.3 & -79.4 &      L  & 7.99   & 8.06   & -0.31 & -0.14  \\ 
\object{F2}   &   8.53 & -0.32  &  35.7 &  4.3 &   ML &   22 &  0.79 &   59.5 &  0.33 &  1.35 & 14.0 & 26.3 &  21.7 &  364.3 &  1.18 &  2.51  &   99.8 &   2.2 &  4.1 &   -8.4 &  -1.1 &    C,X  & 8.32   & 8.67   & -0.36 & -0.22  \\ 
\object{F3}   &   8.76 & -0.37  &  38.5 &  4.1 &   ML &   15 &  0.54 &   38.8 &  0.33 &  0.73 & 11.2 & 26.9 &  15.9 &  410.1 &  1.69 &  4.58  &   80.3 &   6.5 &  4.3 &  -10.6 &  16.3 &    L,X  & 8.64   & 9.03   & -0.41 & -0.31  \\ 
\object{F4}   &   8.16 &  0.22  &  19.8 &  2.9 &   ML &    6 &  0.20 &   10.1 &  0.41 &  0.54 & 13.5 & 24.0 &   4.0 &  396.0 &  1.82 &  5.52  &   22.7 &  16.9 &  5.5 &   29.6 & -20.3 &    L,H  & 8.09   & 8.28   &  0.16 &  0.23  \\ 
\object{F5}   &  10.23 & -0.25  &  12.5 & 13.8 &   ML &   28 &  1.15 &  276.2 &  0.12 &  1.38 & 13.7 & 32.4 & 640.4 & 2318.4 &  1.86 &  0.99  &  120.9 &   2.2 &  5.8 &  -65.4 &  30.0 &    S,X  & 9.95   & 10.43  & -0.42 & -0.07  \\ 
\object{F6}   &  11.07 & -0.39  &  -0.6 & 16.4 &   KF &    6 &  0.18 &   50.1 &  0.06 &  0.30 & 15.0 & 15.0 & 138.4 & 2762.7 &  1.87 &  0.84  &   18.1 &  12.5 &  8.4 & -120.8 & -14.3 &    L,H  & 10.95  & 11.11  & -0.40 & -0.35  \\ 
\object{F7}   &  11.08 & -0.10  &  29.9 &  3.8 &    P &   10 &  0.45 &   30.0 &  0.31 &  1.06 & 12.6 & 15.1 &   9.4 &  312.9 &  0.79 &  1.33  &   43.0 &   4.0 &  4.7 &   10.6 &   0.8 &      S  & 10.97  & 11.3   & -0.13 & -0.04  \\ 
\object{F8}   &  11.75 & -0.10  &  59.7 &  4.8 &   KN &    7 &  0.18 &   15.1 &  0.57 &  1.45 & 15.0 & 15.0 &   3.2 &  212.3 &  0.63 &  1.24  &   28.4 &   1.6 &  3.7 &    6.2 & -72.7 &      C  & 11.73  & 11.8   & -0.15 & -0.07  \\ 
\object{F9}   &  12.48 & -0.22  &  34.7 &  3.3 &   ML &   10 &  0.34 &   19.7 &  0.60 &  1.16 & 15.0 & 16.0 &   2.1 &  108.6 &  0.48 &  1.43  &   46.8 &   4.1 &  5.2 &    5.5 & -53.1 &    L,C  & 12.37  & 12.56  & -0.35 & -0.12  \\ 
\object{F1}   &  12.87 & -0.21  &  35.3 &  2.4 &    P &   35 &  1.50 &   63.0 &  0.65 &  1.47 & 15.0 & 28.0 &  32.3 &  513.0 &  2.51 &  8.11  &  176.1 &   3.0 &  6.0 &   11.3 &  19.0 &    C,X  & 12.69  & 13.33  & -0.37 & -0.05  \\ 
\object{F11}  &  13.29 & -0.32  &  41.1 &  3.5 &   ML &    7 &  0.21 &   12.8 &  0.40 &  0.81 & 13.5 & 15.0 &   4.3 &  334.1 &  1.52 &  4.54  &   31.8 &   5.3 &  5.0 &   -2.0 &  28.2 &      C  & 13.23  & 13.39  & -0.34 & -0.24  \\ 
\object{F12}  &  13.21 &  0.06  &  51.4 &  4.2 &   ML &   10 &  0.42 &   31.0 &  0.33 &  1.77 & 15.0 & 23.3 &   8.6 &  275.9 &  0.88 &  1.85  &   58.2 &   1.5 &  4.3 &   20.4 &  67.1 &      C  & 13.12  & 13.28  &  0.02 &  0.17  \\ 
\object{F13}  &  14.07 & -0.49  &  20.8 &  1.2 &   ML &   25 &  0.85 &   17.3 &  1.02 &  1.46 & 12.1 & 24.2 &   1.7 &  100.6 &  1.13 &  8.42  &   97.7 &   3.7 &  7.2 &   12.6 &  -7.8 &      S  & 13.79  & 14.33  & -0.56 & -0.39  \\ 
\object{F14}  &  14.72 & -0.18  &  39.0 &  3.3 &   ML &   11 &  0.42 &   24.6 &  1.04 &  2.16 & 13.9 & 18.5 &   3.8 &  153.3 &  0.56 &  1.35  &   49.3 &   1.9 &  5.2 &    7.6 & -25.2 &    C,X  & 14.59  & 14.78  & -0.22 & -0.04  \\ 
\object{F15}  &  14.20 & -0.19  &  40.2 &  3.4 &   ML &    7 &  0.31 &   18.3 &  0.36 &  1.15 & 15.0 & 20.5 &   3.4 &  187.9 &  0.75 &  1.95  &   37.9 &   1.5 &  5.1 &    6.5 &   8.5 &      S  & 14.18  & 14.31  & -0.24 & -0.15  \\ 
\object{F16}  &  12.90 &  0.48  &  32.3 &  2.3 &    P &    7 &  0.30 &   12.5 &  0.78 &  1.02 & 15.0 & 25.1 &   1.7 &  133.6 &  0.91 &  4.12  &   41.2 &   4.7 &  6.1 &   39.5 & -59.8 &      S  & 12.83  & 12.99  &  0.35 &  0.55  \\ 
\object{F17}  &  12.79 &  0.36  &  18.9 &  2.1 &   ML &    7 &  0.36 &   13.5 &  0.44 &  0.82 & 15.0 & 21.9 &   1.4 &  102.5 &  0.67 &  2.91  &   46.7 &   1.8 &  6.3 &   33.8 &  70.0 &      S  & 12.7   & 12.86  &  0.32 &  0.52  \\ 
\object{F18}  &  15.05 & -0.66  &  20.5 &  2.0 &    P &   28 &  1.07 &   37.1 &  0.80 &  1.29 & 15.0 & 41.2 &  36.4 &  982.2 &  5.42 &  19.8  &  107.8 &   1.7 &  6.4 &   -2.0 &  27.2 &    X,H  & 14.96  & 15.28  & -0.75 & -0.44  \\ 
\object{F19}  &  19.00 & -0.06  &  61.7 &  4.4 &   ML &    6 &  0.20 &   15.4 &  0.53 &  1.59 & 15.0 & 20.4 &   6.3 &  406.7 &  1.32 &  2.82  &   27.9 &   2.9 &  4.4 &   11.5 &  37.9 &      S  & 18.92  & 19.01  & -0.13 & -0.03  \\ 
\object{F20}  &  20.74 & -0.07  &  57.6 & 11.7 &   ML &   10 &  0.38 &   77.6 &  0.26 &  2.06 & 15.0 & 26.5 &  45.5 &  586.7 &  0.59 &  0.39  &   45.1 &   1.9 &  4.9 &  -13.0 &  68.5 &      C  & 20.66  & 20.78  & -0.16 & -0.01  \\ 
\object{F21}  &  23.10 & -0.39  &  76.9 &  4.6 &    P &    7 &  0.34 &   27.4 &  0.44 &  1.17 & 15.0 & 26.8 &  14.8 &  541.9 &  1.13 &  1.56  &   34.8 &   5.0 &  4.5 &  -15.6 &  13.6 &    L,S  & 23.01  & 23.27  & -0.42 & -0.36  \\ 
\object{F22}  &  23.41 & -0.24  & 103.8 &  5.4 &   ML &    7 &  0.26 &   24.5 &  0.20 &  1.31 & 14.8 & 48.7 &  17.5 &  711.7 &  1.38 &  1.78  &   26.7 &   3.2 &  4.0 &   -7.7 &  52.5 &      S  & 23.35  & 23.46  & -0.34 & -0.21  \\ 
\object{F23}  &  23.37 & -0.12  &  97.5 &  5.9 &    P &    9 &  0.34 &   35.3 &  0.33 &  1.84 & 15.0 & 17.9 &  17.9 &  508.9 &  1.17 &  1.76  &   41.6 &   2.8 &  3.8 &    1.2 & -20.9 &      S  & 23.22  & 23.47  & -0.18 & -0.06  \\ 
\object{F24}  &  23.48 &  0.09  &  83.6 &  4.7 &   ML &    6 &  0.26 &   21.5 &  0.19 &  0.57 & 15.0 & 18.5 &   5.5 &  257.4 &  0.71 &  1.30  &   31.6 &   5.6 &  4.5 &   23.3 &   7.8 &      S  & 23.36  & 23.58  &  0.06 &  0.12  \\ 
\object{F25}  &  24.12 &  0.44  &  94.6 &  7.5 &   ML &    6 &  0.31 &   41.1 &  0.26 &  1.27 & 15.0 & 15.0 &  13.5 &  329.9 &  0.62 &  0.77  &   40.5 &   3.0 &  3.4 &   68.5 & -12.8 &      C  & 24.01  & 24.28  &  0.39 &  0.49  \\ 
\object{F26}  &  24.53 & -0.24  &  98.7 &  5.1 &   ML &    7 &  0.25 &   22.1 &  0.35 &  1.64 & 15.0 & 21.4 &   6.5 &  293.1 &  0.83 &  1.54  &   30.4 &   2.0 &  4.3 &   -6.5 &   8.9 &      C  & 24.44  & 24.58  & -0.28 & -0.21  \\ 
\object{F27}  &  24.79 &  0.10  & 108.7 &  5.6 &   ML &   11 &  0.48 &   47.3 &  0.25 &  1.47 & 11.6 & 34.9 &  26.3 &  555.9 &  1.56 &  2.91  &   69.5 &   2.2 &  4.0 &   24.0 & -60.8 &      S  & 24.73  & 24.86  & -0.03 &  0.18  \\ 
\object{F28}  &  25.30 & -0.22  &  63.3 &  3.9 &   ML &    9 &  0.41 &   27.6 &  0.21 &  0.87 & 10.1 & 24.0 &   8.9 &  321.5 &  1.02 &  2.12  &   45.1 &   3.3 &  5.1 &    2.9 &  16.7 &    L,H  & 25.16  & 25.41  & -0.31 & -0.18  \\ 
\object{F29}  &  25.76 & -0.16  &  93.3 &  4.9 &   ML &   15 &  0.47 &   41.0 &  0.36 &  1.02 & 15.0 & 23.9 &   8.5 &  206.5 &  0.71 &  1.62  &   59.0 &   6.0 &  4.4 &    1.5 & -12.7 &    C,S  & 25.55  & 25.9   & -0.21 & -0.09  \\ 
\object{F30}  &  26.95 &  0.21  &  94.4 &  5.0 &   ML &    6 &  0.29 &   25.3 &  0.22 &  1.12 & 15.0 & 16.6 &   2.3 &   90.7 &  0.23 &  0.38  &   37.5 &   3.1 &  4.5 &   33.8 &  17.1 &      C  & 26.83  & 27.03  &  0.18 &  0.28  \\ 
\object{F31}  &  28.35 &  0.08  &  79.2 &  4.2 &   ML &    8 &  0.17 &   12.8 &  0.64 &  1.29 & 15.0 & 15.0 &   3.4 &  268.6 &  0.79 &  1.52  &   23.8 &   3.1 &  5.0 &   22.9 & -80.2 &      C  & 28.34  & 28.38  &  0.04 &  0.14  \\ 
\object{F32}  &  30.38 & -0.13  & 112.9 &  6.1 &   ML &    6 &  0.26 &   28.3 &  0.23 &  0.99 & 12.8 & 19.8 &  18.2 &  645.3 &  1.24 &  1.57  &   32.5 &   1.6 &  4.3 &    0.4 &  -6.0 &      X  & 30.32  & 30.48  & -0.16 & -0.05  \\ 
\object{F33}  &  32.04 &  0.07  &  95.6 &  5.2 &   ML &    7 &  0.36 &   32.9 &  0.27 &  0.94 & 12.9 & 30.4 &  11.5 &  350.9 &  0.80 &  1.20  &   36.8 &   3.2 &  4.8 &   22.3 &  10.8 &    C,X  & 31.95  & 32.21  &  0.06 &  0.17  \\ 
\object{F34}  &  33.23 &  0.01  & 100.1 &  7.8 &   ML &    8 &  0.32 &   44.0 &  0.17 &  1.44 & 15.0 & 19.8 &  17.1 &  388.7 &  0.48 &  0.38  &   32.9 &   2.0 &  4.7 &   12.9 &  25.6 &      C  & 33.14  & 33.3   & -0.02 &  0.10  \\ 
\object{F35}  &  33.64 & -0.01  & 104.2 &  7.6 &   ML &   11 &  0.40 &   53.1 &  0.28 &  1.01 & 15.0 & 15.6 &  33.3 &  626.6 &  0.96 &  0.97  &   43.7 &   5.3 &  4.7 &   10.4 &  -7.5 &    L,C  & 33.42  & 33.74  & -0.04 &  0.04  \\ 
\end{tabular}
\end{table}
\capstarttrue
\end{turnpage}
\clearpage

\setlength\tabcolsep{4.3pt}
\setcounter{table}{0}
\clearpage
\begin{turnpage}
\capstartfalse
\begin{table}
\caption{Continued.}
\scriptsize
\begin{tabular}{lcccc cccc ccccc ccccc ccccc ccc}
\hline\\
{(1)} 
& {(2)} & {(3)} & {(4)} 
& {(5)} & {(6)} 
& {(7)} 
& {(8)} & {(9)} 
& {(10)} & {(11)} 
& {(12)} & {(13)} 
& {(14)} & {(15)} 
& {(16)} & {(17)}
& {(18)} & {(19)}
& {(20)} & {(21)} & {(22)}
& {(23)}
& {(24)} & {(25)} & {(26)} & {(27)}
\\
{ID} 
& {$l_{\rm wt}$} & {$b_{\rm wt}$} & {$v_{\rm wt}$} 
& {$d$} & {$d_{\rm type}$} 
& {$N_{\rm cl}$} 
& {$L_{\rm deg}$} & {$L_{\rm pc}$} 
& {$\overline{(\frac{\Delta v_i}{\Delta L_i})}$} & {$\sigma_v$} 
& {$T_{\rm min}$} & {$T_{\rm max}$} 
& {Mass} & {$M/L$}
& {$N_{\rm H_2}$} & {$n_{\rm H_2}$} 
& {$f_A$} & {$f_L$} 
& {$R_{\rm gc}$} & {$z$} & {$\theta$}
& {Morph.} 
& {$l_{\rm min}$} & {$l_{\rm max}$}
& {$b_{\rm min}$} & {$b_{\rm max}$}\\
\hline \\
\object{F36}  &  34.26 &  0.14 &  57.7 &  1.7 &   ML &   15 &  0.69 &   20.9 &  1.09 &  1.13 &  5.5 & 44.5 &  16.2 &  775.6  &  4.52 & 17.41  &   72.3 &   2.8 &  7.0 &   26.4 &  27.8 &    L,X  & 34.09  & 34.46  &  0.02  &  0.25 \\
\object{F37}  &  37.39 & -0.07 &  57.2 &  9.6 &   ML &    7 &  0.26 &   44.0 &  0.15 &  1.06 & 15.0 & 25.9 &  14.0 &  317.4  &  0.51 &  0.54  &   31.6 &   3.0 &  5.9 &   -1.8 &   3.4 &      C  & 37.30  & 37.48  & -0.10  & -0.05 \\
\object{F38}  &  41.18 & -0.21 &  59.3 &  8.6 &   ML &    7 &  0.27 &   40.9 &  0.18 &  1.33 & 15.0 & 40.8 &  18.9 &  461.1  &  0.60 &  0.52  &   28.7 &   5.9 &  6.0 &  -18.5 &   9.2 &      S  & 41.10  & 41.33  & -0.24  & -0.17 \\
\object{F39}  &  45.46 &  0.07 &  57.7 &  8.4 &    P &    7 &  0.26 &   37.6 &  0.20 &  1.17 & 15.0 & 15.0 &  39.3 & 1045.7  &  1.62 &  1.65  &   33.2 &   1.7 &  6.5 &   24.9 &  22.9 &      X  & 45.39  & 45.54  &  0.01  &  0.14 \\
\object{F40}  &  45.82 & -0.32 &  60.1 &  7.1 &   ML &   12 &  0.59 &   73.4 &  0.21 &  2.35 & 15.0 & 20.4 &  13.0 &  177.7  &  0.37 &  0.51  &   74.7 &   2.8 &  6.1 &  -23.6 & -31.9 &    L,X  & 45.68  & 45.94  & -0.40  & -0.22 \\
\object{F41}  &  48.97 & -0.29 &  68.6 &  4.9 &   ML &   10 &  0.37 &   31.5 &  0.69 &  2.12 & 15.0 & 30.2 &  15.2 &  483.5  &  1.15 &  1.82  &   43.1 &   2.0 &  6.3 &   -5.2 &   0.7 &      X  & 48.91  & 49.10  & -0.35  & -0.25 \\
\object{F42}  &  49.49 & -0.37 &  60.0 &  5.4 &   ML &    6 &  0.24 &   23.2 &  0.33 &  1.56 & 15.0 & 23.1 & 199.9 & 8624.7  & 16.94 &  22.0  &   24.1 &   3.5 &  6.3 &  -15.9 & -55.7 &      C  & 49.47  & 49.57  & -0.40  & -0.29 \\
\object{F43}  &  78.02 &  0.60 &  -2.9 &  3.7 &   KF &    6 &  0.17 &   10.8 &  0.34 &  1.20 & 15.0 & 15.0 &   2.6 &  244.3  &  0.89 &  2.15  &   20.1 &   1.9 &  8.4 &   63.9 &  32.5 &      C  & 77.99  & 78.06  &  0.56  &  0.62 \\
\object{F44}  &  77.91 & -1.17 &  -0.6 &  3.2 &   KF &    7 &  0.20 &   11.4 &  0.55 &  0.83 & 15.0 & 15.0 &   1.9 &  166.4  &  0.71 &  2.01  &   25.6 &   7.7 &  8.3 &  -39.9 &  53.4 &    L,C  & 77.84  & 77.95  & -1.27  & -1.14 \\
\object{F45}  &  81.25 &  0.98 &  14.2 &  1.3 &   KN &   16 &  0.62 &   13.7 &  1.67 &  1.81 &  9.1 & 26.1 &   0.7 &   53.5  &  0.49 &  2.96  &   76.4 &   2.9 &  8.2 &   47.1 &  87.2 &    S,X  & 81.15  & 81.30  &  0.80  &  1.12 \\
\object{F46}  &  81.71 &  0.56 &  -2.2 &  1.5 &   ML &   13 &  0.56 &   14.6 &  0.94 &  1.76 & 10.2 & 29.6 &   5.9 &  406.9  &  3.17 & 16.32  &   61.8 &   7.9 &  8.3 &   40.2 &  56.2 &    L,C  & 81.56  & 81.80  &  0.28  &  0.66 \\
\object{F47}  &  81.51 &  0.04 &  -5.4 &  3.2 &   KN &   13 &  0.49 &   27.5 &  0.31 &  0.96 & 11.1 & 24.7 &   5.5 &  199.6  &  0.61 &  1.25  &   46.5 &   4.3 &  8.5 &   28.6 &  32.8 &    S,X  & 81.40  & 81.69  & -0.06  &  0.15 \\
\object{F48}  &  81.03 & -0.13 &  -3.4 &  3.1 &   KN &    7 &  0.35 &   19.2 &  0.20 &  1.26 &  9.2 & 23.3 &   9.3 &  485.3  &  1.51 &  3.09  &   33.4 &   2.4 &  8.4 &   19.2 &   2.0 &      C  & 80.94  & 81.18  & -0.15  & -0.06 \\
\object{F49}  & 110.06 & -0.11 & -51.8 &  4.4 &   ML &    9 &  0.42 &   32.3 &  0.20 &  0.67 & 15.0 & 15.0 &   7.1 &  218.8  &  0.67 &  1.37  &   55.9 &   2.9 & 10.7 &   24.4 &  82.3 &      S  & 110.00 & 110.10 & -0.28  & -0.03 \\
\object{F50}  & 111.53 &  0.78 & -56.8 &  2.6 &    P &   21 &  0.97 &   45.0 &  0.50 &  2.05 &  4.8 & 33.5 &  23.7 &  526.0  &  1.87 &  4.39  &   94.9 &   1.6 &  9.6 &   66.0 &   2.5 &      X  & 111.37 & 111.63 &  0.66  &  0.84 \\
\object{F51}  & 111.82 &  0.68 & -51.8 &  4.4 &   ML &   21 &  0.87 &   66.5 &  0.28 &  1.35 &  6.2 & 19.6 &  42.4 &  637.5  &  1.34 &  1.86  &   73.3 &   1.8 & 10.8 &   85.6 &  80.1 &    C,X  & 111.70 & 111.95 &  0.46  &  0.86 \\
\object{F52}  & 111.63 &  0.37 & -56.1 &  4.6 &   ML &    6 &  0.24 &   19.3 &  0.42 &  0.74 & 15.0 & 16.5 &   6.9 &  356.6  &  0.86 &  1.38  &   26.1 &   2.5 & 10.9 &   64.2 & -12.9 &      H  & 111.62 & 111.76 &  0.33  &  0.41 \\
\object{F53}  & 134.21 &  0.75 & -48.4 &  4.1 &   ML &    8 &  0.31 &   22.1 &  0.31 &  1.50 &  7.0 & 15.0 &  49.6 & 2242.2  &  4.91 &  7.11  &   27.3 &   2.8 & 11.6 &   90.5 &  85.0 &      C  & 134.20 & 134.28 &  0.64  &  0.86 \\
\object{F54}  & 192.62 & -0.03 &   8.0 &  1.6 &   ML &   10 &  0.50 &   13.8 &  0.93 &  1.08 &  9.2 & 42.9 &   2.2 &  156.0  &  1.18 &  5.85  &   55.8 &   3.2 &  9.9 &   30.1 &  57.1 &      C  & 192.58 & 192.82 & -0.16  &  0.13 \\
\hline \\
Min  &   8.01 & -1.17 & -56.8 &  1.2 &      &    6 &  0.17 &   10.1 &  0.06 &  0.30 &  4.8 & 15.0 &   0.7 &   53.5  &  0.23 &  0.38  &   18.1 &   1.5 &  3.4 & -120.8 & -80.2 \\
Max  & 192.62 &  0.98 & 112.9 & 16.4 &      &   35 &  1.50 &  276.2 &  1.67 &  2.35 & 15.0 & 48.7 & 640.4 & 8624.7  & 16.94 &  22.0  &  176.1 &  16.9 & 11.6 &   90.5 &  87.2 \\
Med  &  25.04 & -0.10 &  40.6 &  4.4 &      &    8 &  0.35 &   27.5 &  0.33 &  1.26 & 15.0 & 23.2 &  10.5 &  360.5  &  0.94 &  1.77  &   41.4 &   3.0 &  5.4 &   10.9 &   9.1 \\
Mean &  40.95 & -0.02 &  41.8 &  5.1 &      &   10 &  0.43 &   35.7 &  0.43 &  1.27 & 13.3 & 23.9 &  31.2 &  644.1  &  1.57 &  3.53  &   50.0 &   3.7 &  6.1 &   11.5 &   9.8 \\
Std  &  38.95 &  0.38 &  45.2 &  3.1 &      &    6 &  0.27 &   37.3 &  0.31 &  0.43 &  2.8 &  8.5 &  90.5 & 1224.6  &  2.38 &   4.8  &   29.7 &   2.7 &  2.1 &   35.5 &  42.0 \\
$S$  &   1.78 &  0.32 &  -0.5 &  1.7 &      &    1 &  1.91 &    5.1 &  1.79 &  0.35 & -1.7 &  1.1 &   5.9 &    5.4  &  0.53 &  0.26  &    2.0 &   2.9 &  1.0 &   -0.7 &  -0.2 \\
$K$  &   3.11 &  1.11 &  -0.3 &  3.0 &      &    3 &  3.73 &   30.3 &  3.69 & -0.01 &  1.8 &  0.7 &  36.7 &   32.0  &  3.08 &  0.62  &    4.7 &  10.4 &  0.0 &    2.8 &  -0.3 \\
\hline
\end{tabular}
\tablecomments{A brief description of the columns (for a detailed description including uncertainties see \autoref{sec:para}):
Column (1) assigned ID.
Col. (2--4) flux weighted longitude, latitude (in degree), and LSR velocity (\kms).
Col. (5--6) distance (kpc) and its type.
Col. (7) number of clumps in the filament.
Col. (8--9) length of the filament in degree and pc.
Col. (10) velocity gradient (\kms\,pc$^{-1}$).
Col. (11) dispersion of the central velocity (\kms) of all the clumps in the filament.
Col. (12--13) minimum and maximum temperature of the clumps in the filament.
Col. (14-15) filament mass (in unit $10^3$\msun) and linear mass density (\msun/pc).
Col. (16--17) estimated H$_2$ column density ($10^{22}$ \cms) and volume density ($10^3$ \cmc) of the filament.
Col. (18--19) aspect ratio and linearity.
Col. (20) Galactocentric radius (kpc).
Col. (21) vertical distance (pc) to the physical Galactic mid-plane.
Col. (22) orientation angle (degree) between the filament's long axis and the physical Galactic mid-plane. Positive/negative angle means Galactic latitude increases/decreases with increasing longitude.
Col. (23) Morphology class.
Col. (24--27) Galactic coordinate boundary of the filament (degree).
}
\end{table}
\capstarttrue
\end{turnpage}
\clearpage


\bibliographystyle{yahapj}
\bibliography{my.astro}

\end{document}